\documentclass[a4paper,12pt,aip]{article}
\usepackage[utf8x]{inputenc}
\usepackage[T1]{fontenc}
\usepackage{amsmath,amsfonts,amsbsy, amssymb}
\usepackage{color,xcolor}
\usepackage{graphicx}
\usepackage{isotope}
\usepackage{a4wide}
\usepackage{caption}
\usepackage{cite}
\input ArtNouv.fd
\newcommand*\initfamily{\usefont{U}{ArtNouv}{xl}{n}}
\usepackage{authblk}
\usepackage{multido}
\usepackage{palatino}
\usepackage[pdftex,colorlinks=true, linkcolor=blue]{hyperref}
\hypersetup{pdfborder=0 0 1}
\def\vec#1{\boldsymbol{#1}}
\def\piz      {\ensuremath{\pi^0}}
\def\pip      {\ensuremath{\pi^+}}
\def\pim      {\ensuremath{\pi^-}}

\def\Kp    {\ensuremath{ {K}^+}}
\def\Km    {\ensuremath{ {K}^-}}

\def\Ks    {\ensuremath{ {K_s}}}
\def\Kl    {\ensuremath{ {K_l}}} 
\newcommand{\SLJ}[3]    {\relax\ensuremath{{}^{#1}{{\mathrm #2}}_{#3}}}
\newcommand{\ISLJ}[4]   {\relax\ensuremath{{}^{#1,#2}{{\mathrm #3}}_{#4}}}

\newcommand{\islj}{\relax\ensuremath{{}^{2I+1,2S+1}L_J}}
\newcommand{\slj}{\relax\ensuremath{{}^{2S+1}L_J}}

\def\NNb              {\ensuremath{\bar N N}}

\DeclareMathOperator{\IM}{Im}
\DeclareMathOperator{\RE}{Re}
\usepackage{appendix}
\newcommand{\RomanNumeralCaps}[1]
    {\MakeUppercase{\romannumeral #1}}
\begin{document}
\author[1]{Jean-Marc Richard}
\affil[1]{Institut de Physique des 2 Infinis de Lyon \& Universit\'e de Lyon\\
UCBL, CNRS-IN2P3\\
4, rue Enrico Fermi, Villeurbanne, France}
\title{\Huge{\initfamily ANTIPROTON PHYSICS}\footnote{To appear in the collection of review articles \emph{The long-lasting quest for nuclear interactions: the past, the present and the future} of the journal Frontiers in Physics, eds. Laura Eliss Marcucci and Rupert Machleidt }
\date{December 16, 2019}}
\maketitle
\begin{abstract}
 We review the physics of low-energy antiprotons, and its link with the nuclear forces. This includes: antinucleon scattering on nucleons and nuclei, antiprotonic atoms and  antinucleon-nucleon  annihilation into mesons. 
\end{abstract}
\tableofcontents
\section{A brief history}\label{se:disco}
In 1932, the positron, the antiparticle of the electron,  was discovered in cosmic rays and confirmed in the $\beta^+$ decay of some radioactive nuclei. See, e.g., the Nobel lecture by Anderson \cite{Brink:204994}. It was then reasonably  anticipated that the proton also has an antiparticle, the \emph{antiproton}.\footnote{The only doubt came from the magnetic moment of the proton, which is not what is expected for a particle obeying the Dirac equation.}\@ It was also suspected that the antiproton would be more difficult to  produce and detect than positrons in cosmic rays. The Bevatron project (BeV, i.e., billion of electron-volts, was then a standard denomination for what is now GeV) was launched at Berkeley to reach an energy high-enough to produce antiprotons through the reaction 
\begin{equation}
 p+A\to p+A+\bar p+p~,
\end{equation}
where $A$ denotes the target nucleus. For $A=p$, this is a standard exercise in relativistic kinematics to demonstrate that the kinetic energy of the incoming proton should be higher than $6\,m$, where $m$ is the proton mass, and $c=1$. This threshold decreases if the target is more massive.  The Bevatron was completed in 1954, and the antiproton was discovered in 1955 by a team lead by Chamberlain and Segrè, who were awarded the Nobel prize in 1959.\footnote{The other collaborators were acknowledged in the Nobel lectures, but nevertheless Piccioni sued Chamberlain and Segrè in a court of California, which dismissed the suit on procedural grounds}

Shortly after the antiproton, the antineutron, $\bar n$,  was also discovered at Berkeley, and up to now, for any new elementary particle, the corresponding antiparticle has also been found. The discovery of the first anti-atom was well advertised \cite{1996PhLB..368..251B}, but this was not the case for the earlier observation of the first antinucleus, antideuterium, because of a controversy between an European team \cite{1965NCimS..39...10M}  and its US competitors \cite{Dorfan:1965uf}. In  experiments at very high energy, in particular collisions of heavy ions at STAR (Brookhaven) and ALICE (CERN), one routinely produces light antinuclei and even anti-hypernuclei (in which an antinucleon is replaced by an antihyperon $\bar\Lambda$)~\cite{Ma:2013voa,Adam:2019phl,Donigus:2019qvq}.

The matter-antimatter symmetry is almost perfect, except for a slight violation in the sector of weak interactions, which is nearly exactly compensated by a simultaneous violation of the left-right symmetry, i.e., the product $PC$ of parity $P$ and charge-conjugation $C$ is only very marginally violated. Up to now, there is no indication of any violation of the product $CPT$, where $T$ is the time-reversal operator: this implies that the proton and antiproton have the same mass, a property now checked to less than $10^{-9}$ \cite{Hori:2011zz}. 

% It should be noted that that beyond antiparticles, antimatter has been produced. The first evidence for antihydrogen by W. Oelert et al.  was much advertised, and the properties of antihydrogen are now measured in great detail at CERN. Previously, antideuterium $(\bar d)$ was seen by Zichichi et al. \cite{1965NCimA..63...10M} in Europe and by another team in the US 
% %\cite{Dorfan:1965uf}. 
% In high-energy collision, one routinely produces light antinuclei and even anti-hypernuclei (in which an antinucleon is replaced by an antihyperon $\bar\Lambda$.  In devices producing antiprotons, a non-negligible fraction of $\bar d$, which could be cooled and stored: the intensity would be sufficient to perform strong-interaction measurements.
% 

Many experiments have been carried out with low energy antiprotons, in particular at Brookhaven and CERN in the 60s and 70s, with interesting results, in particular for the physics of the mesons produced by annihilation. However, in these early experiments, the antiprotons were part of secondary beams containing many negatively-charged pions and kaons, and with a wide momentum spread. 

In the 70s, Simon van der Meer, and his colleagues at CERN and elsewhere imagined and developed the method of \emph{stochastic cooling} \cite{Mohl:2013ri}, which produces antiproton beams of high purity, sharp momentum resolution, and much higher intensity than in the previous devices.  CERN transformed the fixed-target accelerator SpS into a proton-antiproton collider, Sp$\bar{\rm p}$S, with the striking achievement of the discovery of the $W^\pm$ and $Z^0$, the intermediate bosons of the electro-weak interaction.  A similar scheme was later adopted at Fermilab with higher energy and intensity, leading to many results, among which the discovery of the top quark.  

As a side product of the experiments at the Sp$\bar{\rm p}$S program, CERN built a low-energy facility, LEAR  (Low-Energy Antiproton Ring) which operated from 1982 to 1996, and hosted several experiments on which we shall come back later.  Today, the antiproton source of CERN is mainly devoted to experiments dealing with atomic physics and fundamental symmetries.  
In spite of several interesting proposals, no low-energy extension of the antiproton program was built at Fermilab.  

As for the intermediate energies, at the beginning of the CERN cooled-antiproton program, a $\bar p$ beam was sent in the ISR accelerator to hit a thin hydrogen target. The experiment R704 got sharp peaks corresponding to some charmonium states, and in particular a first indication of the --then missing -- P-wave singlet state $h_c$ \cite{Baglin:1985xa}. But ISR was  to be closed, and in spite of a few more days of run, R704 was interrupted. The team moved to Fermilab,  and charmonium physics with antiprotons was resumed  with  antiproton-proton collisions  arranged in the accumulation device (experiments E760-E835) \cite{Patrignani:2004nf}.

Today, the techniques of production of sharp antiproton beams is well undercontrol. There are projects to perform strong-interaction physics with antiprotons at FAIR (Darmstadt) \cite{Augustin:2008gy} and JPARC in Japan \cite{Kumano:2015gna}. In the 80s, an ambitious extension of LEAR at higher energies, SuperLEAR \cite{Dalpiaz:1987qk}, was proposed  by L.~Montanet et al.,  but was not approved by the CERN management. A major focus of SuperLEAR was charm physics. But more than 30 years later, this physics has been largely unveiled by beauty factories and high-energy hadron colliders. 

Presently, the only running  source of cooled antiprotons is the very low energy AD at CERN (Antiproton Decelerator) and its extension ELENA (Extra Low ENergy Antiproton) with the purpose of doing atomic-physics and high-precision tests of fundamental symmetries. Some further decelerating devices are envisaged for the gravitation experiments \cite{Husson:2019nzl}.  Of course, standard secondary antiproton beams are routinely produced, e.g., at KEK in Japan. 

Note also that in devices making antiproton beams, a non-negligible fraction of antideuterium is produced, which could be cooled and stored. The intensity would be sufficient to perform strong-interaction measurements, but there is not yet any proposal for an experiment with an antideuterium beam.

We shall discuss along this review  many results obtained at LEAR and elsewhere. Already the measurements made at Berkeley during the weeks following the discovery of the antiproton were remarkable. After more than 60 years, we realize today that  they gave keys to the modern understanding of hadrons, but the correct interpretation was too far from the current wisdom of the 50s. Indeed, from the work by Fermi and Yang, on which more later, it was realized that one-pion exchange constitutes the long-range part of the antinucleon-nucleon interaction. The simplest model, just before the discovery of the antiproton, would be one-pion exchange supplemented by a very short-range annihilation. This would imply for the charge-exchange ($\bar p p\to \bar n n$), elastic ($\bar p p\to \bar p p$) and annihilation ($\bar p p\to \text{mesons}$) cross-sections a hierarchy
\begin{equation}
 \sigma_\text{ce}>\sigma_\text{el}>\sigma_\text{an}~,
\end{equation}
the first inequality resulting from straightforward isospin algebra. What was observed at Berkeley is just the opposite! And it took us years to admit  and understand this pattern, which is a consequence of the composite character of the nucleon and antinucleon. 

The era of LEAR and Sp$\bar{\rm p}$S at CERN, and then the large $\bar p p$ collider of Fermilab will  certainly be reminded as the culmination of antiproton physics. At very high energy, the trend is now more on $pp$ rather than $\bar p p$ collision, due to the higher intensity of proton beams. Certainly very-low energy experiments will remain on the floor to probe the fundamental symmetries with higher and higher precision. The question is open on whether antiproton beams will be used for hadron physics, a field where electron beams and flavor factories already provide much information. 

Of course, the role of antimatter in astrophysics is of the highest importance. Antiprotons and even antinuclei are seen  in  high-energy cosmic rays. 
The question is to estimate how many antinuclei are expected to be produced by standard cosmic rays, to estimate the rate of primary antinuclei. See, e.g., \cite{Chardonnet:1997dv, Aguilar:2016kjl}. Some years ago, cosmological models were built  \cite{Aly:1974yq}  in which the same amount of matter and antimatter was created, with a separation of zones of matter and zones of antimatter. In modern cosmology, it is assumed that an asymmetry prevailed, so that, after annihilation, some matter survived. 

This review is mainly devoted to the low-energy experiments with antinucleons. Needless to say that the literature is abundant, starting with  dedicated workshops \cite{Gastaldi:99763,Gastaldi:106464, Charalambous:1987ny,Amsler:112013,Amsler:237631,Pinsky:1986nd} and schools~\cite{Bloch:242185,Gastaldi:1989hr,Bradamante:1990hp,Landua:1991tn}.

Because of the lack of space, some important subjects will not be discussed, in particular the ones related to fundamental symmetry: the inertial mass of the antiproton, its charge and magnetic moment, in comparison with the values for the proton; the detailed comparison of hydrogen and antihydrogen atoms; the gravitational properties of neutral atoms such as antihydrogen, etc.  We will mention only very briefly, in the section on antiprotonic atoms, the dramatically precise atomic physics made with  the antiprotonic Helium. 
\section{From the nucleon-nucleon to the antinucleon-nucleon interaction}
\label{se:NN-to-NbarN}

In this section we outline the general theoretical framework: how to extrapolate our information on the nuclear forces to the antinucleon-nucleon system. We present the basics of the well-known $G$-parity rule, with a remark about the definition of antiparticle states. 

\boldmath\subsection{The $G$-parity rule}\unboldmath
In QED, the $e^-e^-\to e^-e^-$  and  $e^+e^-\to e^+e^-$ amplitudes are related by crossing, but there is a long way from  the region $\{s>4\,m_e^2,t<0\}$ to the one with$\{s<0, t>4\,m_e^2$ to  attempt a reliable analytic extrapolation. Here, $m_e$ is the electron mass and $s$ and $t$ the usual Mandelstam variables.%
\footnote{For a reaction $1+2\to 3+4$, the Mandestam variables are given in terms of the energy-momentum quadrivectors as $s=(\tilde p_1+\tilde p_2)^2$, $t=(\tilde p_3-\tilde p_1)^2$ and $u=(\tilde p_4-\tilde p_1)^2$.}\@
A  more useful approach consists of comparing both reactions for  the same values of $s$ and $t$.  The  $e^-e^-\to e^-e^-$ amplitude can be decomposed  into a even and odd part according to the $C$-conjugation in the $t$-channel, say
\begin{equation}
 \mathcal{M}(e^-e^-\to e^-e^-)=\mathcal{M}_\text{even}+\mathcal{M}_\text{odd}~,
\end{equation}
and the  $e^+e^-\to e^+e^-$ amplitude for the same energy and transfer is given by
\begin{equation}
 \mathcal{M}(e^+e^-\to e^+e^-)=\mathcal{M}_\text{even}-\mathcal{M}_\text{odd}~.
\end{equation}
The first term contains the exchange of an even number of photons,  and the last one the exchange of an odd number. At lowest order, one retrieves the sign flip of the Coulomb potential.  This rule remains valid to link $pp\to pp$ and $\bar p p\to \bar p p$ amplitudes: the exchange of a $\pi^0$ with charge conjugaison $C=+1$, is the same for both reactions, while the exchange of an $\omega$ meson ($C=-1$) flips sign. 

In \cite{Fermi:1959sa}, Fermi and Yang astutely combined this $C$-conjugation rule with isospin symmetry, allowing to include the exchange of charged mesons, as in the charge-exchange processes. Instead of comparing $pp\to pp$ to $\bar p p\to \bar p p$ or $np\to np$ to $\bar n p\to \bar n p$, the $G$-parity rule relates amplitudes of given isospin $I$.  More precisely, if the nucleon-nucleon amplitude is decomposed as 
\begin{equation}
\mathcal{M}^I(NN)=\mathcal{M}_\text{G=+1}+\mathcal{M}_\text{G=-1}~,
\end{equation}
according to the $G$-odd (pion, omega, \dots) or $G$-even ($\rho$, \dots) in the $t$-channel, then its $\bar N N$ counterpart reads
\begin{equation}\label{eq:G-parity}
 \mathcal{M}^I(\bar NN)=\mathcal{M}_\text{G=+1}-\mathcal{M}_\text{G=-1}~.
\end{equation}
Note that there is sometimes some confusion between the $C$-conjugation  and the $G$-parity rules, especially because there are two ways of defining the isospin doublet $\{\bar n, \bar p\}$. See appendix \ref{app:A}.

In current models of $NN$, the pion-exchange tail, the attraction due to isoscalar two-pion exchange, and  the spin-dependent part of the $\rho$ exchange are rather well identified, and thus can be rather safely transcribed in the $\bar N N$ sector.  Other terms, such as the central repulsion attributed to $\omega$-exchange, might contain contributions carrying the opposite $G$-parity,  hidden in the effective adjustment of the couplings. Thus the translation towards $\bar N N$ might be biased.
\subsection{Properties of the long-range interaction}
Some important consequences of the $G$-parity rule have been identified.
First, the moderate attraction observed in $NN$, due to a partial cancellation of $\sigma$ (or, say, the scalar-isoscalar part of two-pion exchange) and $\omega$-exchanges, becomes a coherent attraction once $\omega$-exchange flips sign. This led Fermi and Yang to question whether the mesons could be interpreted as bound states of a nucleon and an antinucleon. This idea  has been regularly revisited, in particular at the time of bootstrap  \cite{Jacob:105219}.  As stressed, e.g., in \cite{Ball:1965sa}, this approach hardly accounts for the observed degeneracy of $I=0$ and $I=1$ mesons (for instance $\omega$ and $\rho$ having about the same mass). 

In the 70s, Shapiro et al., and others, suggested that baryon-antibaryon bound states were associated with new types of hadrons, with the name \emph{baryonium}, or \emph{quasi-deuteron} \cite{Shapiro:1978wi,Buck:1977rt,Dover:1979zj}. Similar speculations were made later for other hadron-hadron systems, for instance $D\bar D{}^*$, where $D$ is a charmed meson $(\bar c q)$ of spin 0 and $\bar D{}^*$ an anticharmed meson $(\bar c q)$ of spin 1 \cite{Tornqvist:1993ng}.  Some candidates for baryonium were found in the late 70s, interpreted either as quasi-nuclear $\bar N N$ states \`a la Shapiro, or as exotic states in the quark model, and motivated the construction of the LEAR facility at CERN. Unfortunately, the baryonium states were not confirmed. 

Another consequence of the  $G$-parity rule is a dramatic change of the spin dependence of the interaction. At very low energy, the nucleon-nucleon interaction is dominated by the spin-spin and tensor contributions of the one-pion exchange. However, when the energy increases, or, equivalently, when one explores shorter distances, the main pattern is a pronounced spin-orbit interaction. It results from a coherent sum of the contributions of vector mesons and scalar mesons.\footnote{The origin is different, for vector mesons, this is a genuine spin-orbit effect, for scalar mesons, this is a consequence of Thomas precession, but the effect is the same in practice}\@ The tensor component of the $NN$ interaction is known to play a crucial role: in most models, the $\SLJ{1}{S}{0}$ potential is stronger than the  $\SLJ{3}{S}{1}$ one, but in this  latter partial wave,\footnote{The notation  is $\slj$, as there is a single choice of isospin, and it will become $\islj$ for $\bar N N$} the attraction is reinforced by S-D mixing. However, the effect of the tensor force remains moderate, with a percentage of D wave of about $5\%$ for the deuteron. 

In the case of the $\bar NN$ interaction, the most striking coherence occurs in the tensor potential, especially in the case of isospin $I=0$ \cite{Dover:1978br}. A scenario with dominant tensor forces is somewhat unusual, and leads to unexpected consequences, in particular a relaxation of the familiar hierarchy based on the hight of the centrifugal barrier. For instance, if one calculates the spectrum of bound states from the real part of the $\bar N N$ interaction, the ground-state is $\ISLJ13P0$, and next a coherent superposition of $\ISLJ13S1$ and $\ISLJ13D1$, and so on.  In a scattering process, there is no polarization if the tensor component is treated to first order,  but polarization shows up at higher order. Thus, one needs more than polarization measurements\footnote{Actually more than spin measurements along the normal $\hat n$ to the scattering plane, such as the analyzing power $A_n$ or the transfer $D_{nn}$ or normal polarization} to  distinguish the  dynamics with a moderate spin-orbit component from the dynamics with  a very strong tensor component.
\begin{subappendices}
\subsection{Appendix: Isospin conventions}
\label{app:A}
There are two possible conventions for writing the isospin states of antinucleons \cite{Martin:102663}. 

The natural choice is based on the charge conjugation operator $C$, namely $|\bar p\rangle\strut_c=C\,|p\rangle$ and $|\bar n\rangle\strut_c=C\,|n\rangle$. However, it transforms the 2 representation of SU(2)  into a $\bar 2$ which does not couple with the usual Clebsch-Gordan coefficients. For instance, the isospin $I=0$ state of $\bar NN$ reads in this convention
\begin{equation}
 |I=0\rangle=\frac{|\bar p p\rangle\strut_c+|\bar n n\rangle\strut_c}{\sqrt2}~,
\end{equation}
which anticipates the formula for an SU(3) singlet, 
\begin{equation}
 |0\rangle=\frac{|\bar u u\rangle+|\bar d d\rangle+|\bar s s\rangle}{\sqrt3}~,
\end{equation}

However, the $\bar 2$ representation of SU(2)  is equivalent to the 2 one, and it turns out convenient to perform the corresponding rotation, that is to say, define the states by the $G$-parity operator, namely (without subscript) $|\bar p\rangle=G\,|n\rangle$ and $|\bar n\rangle=G\,|p\rangle$. With this convention, the isospin singlet is written as 
\begin{equation}
 |I=0\rangle=\frac{|\bar n n\rangle-|\bar p p\rangle}{\sqrt2}~,
\end{equation}
\end{subappendices}
\section{Baryonium}\label{se:baryonium}
The occurrence of baryonium candidates in antiproton-induced reactions was a major subject  of discussion in the late 70s and in the 80s and the main motivation to build new antiprotons beams and new detectors.   The name ``baryonium'' suggests a baryon-antibaryon structure, as in the quasi-nuclear models.  More generally  ``baryonium'' denotes mesons that are preferentially coupled to the baryon-antibaryon channel, independently of any prejudice about their internal structure. 

Nowadays, baryonium is almost dead, but interestingly, some of the innovative concepts and some unjustified approximations developed for baryonium are re-used in the current discussions about the new hidden-charm mesons $XYZ$ and other exotic hadrons \cite{Brambilla:2019esw}.
\subsection{Experimental candidates for baryonium}
For an early review on baryonium, see \cite{Montanet:1980te}. For an update, see the Particle Data Group \cite{Tanabashi:2018oca}.  In short: peaks have been seen  in the integrated cross sections, or in the angular distribution (differential cross section) at given angle, or in some specific annihilation rates as a function of the energy.  The most famous candidate was the S(1932), seen in several experiments  \cite{Montanet:1980te}. The most striking candidate was the peak of mass  2.95\,GeV$/c^2$ seen in $\bar p p\pi^-$ \cite{Evangelista:1977ni}, with some weaker evidence for peaks at 2.0 and 2.2\,GeV$/c^2$ in the $\bar p p$ subsystem, suggesting a sequential decay $\mathcal{B}^-\to \mathcal{B}+\pi^-$, where $\mathcal{B}$ denotes a baryonium. 
Peaks were also seen in the inclusive photon and pion spectra of the annihilations  $\bar p p\to \gamma X$ and $\bar p p\to \pi X$ at rest. 

None of the experiments carried out at LEAR confirmed the existence of such peaks.  However, some enhancements have been seen more recently in the $\bar p p$ mass distribution of the decay of heavy particles, such as $J/\psi\to \gamma \bar p p$, $B\to K  \bar p p$ or $B\to D  \bar p p$, see  \cite{BESIII:2011aa} and  the notice on non $\bar q q$ mesons in \cite{Tanabashi:2018oca}.  There is a debate about whether they correspond a baryonium states or just  reveal a strong $\bar p p$ interaction in the final state.  See, e.g., the discussion in \cite{Loiseau:2005cv,Wycech:2012zza,Kang:2015yka}.  Also, as stressed by Amsler  \cite{Amsler:2019ytk}, the $f_2(1565)$ \cite{f21565} is seen only in annihilation experiments, and thus could be a type of baryonium, $\ISLJ13P2-\ISLJ13F2$ in the quasi-nuclear models. 
\subsection{The quasi-nuclear model}
Today, it is named ``molecular'' approach. The observation that the real part of the $\bar N N$ interaction is more attractive than its $NN$ counterpart led Shapiro et al.~\cite{Shapiro:1978wi},  Dover et al.~\cite{Buck:1977rt}, and others, to predict the existence of deuteron-like $\bar N N$ bound states and resonances.  Due to the pronounced spin-isospin dependence of the $\bar N N$ interaction, states with isospin $I=0$ and natural parity were privileged in the predictions. The least one should say is that the role of annihilation was underestimated in most early studies. Attempts to include annihilation in the spectral problem have shown, indeed, that most structures created by the real potential are washed out when the absorptive part is switched on~\cite{Myhrer:1976by}. 
% Possible exceptions: 1) states with high angular momentum. 2) Scenario in which annihilation proceeds by baryonium plus pion, so that the lowest baryonium might be rather narrow???
%
\subsection{Duality}
Duality is a very interesting concept developed in the 60s. For our purpose, the most important aspect is that in a hadronic reaction $a+b\to c+d$, there is an equivalence between the $t$-channel dynamics, i.e., the exchanges schematically summarized as $\sum _i\, a+\bar c\to X_i\to \bar b+d$, and the low-energy resonances $\sum_j \, a+b\to Y_j \to c+d$. In practice, one approach is usually more efficient than the other, but a warning was set by duality against empirical superpositions of $t$-channel and $s$-channel contributions.  For instance, $\bar K N$ scattering with strangeness $S=-1$ benefits  the hyperons as $s$-channel resonances, and one also observes a coherent  effect of the exchanged mesons. On the other hand, $K N$ is exotic, and, indeed, has a much smaller cross-section. In $KN$, there should be destructive interferences among the $t$-channel exchanges. 

\begin{figure}[ht!]
 \centering
 \includegraphics[width=.2\textwidth]{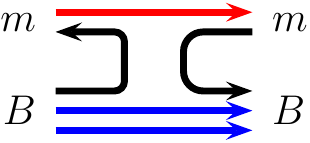}
 % duality-fig7.pdf: 0x0 pixel, 300dpi, 0.00x0.00 cm, bb=
 \caption{Duality diagram for non-exotic meson-baryon scattering}
 \label{fig:dual7}
\end{figure}

\begin{figure}[ht!]
 \centering
 \includegraphics[width=.2\textwidth]{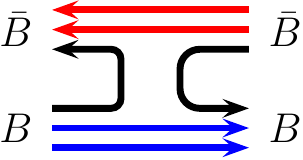}
 % duality-fig7.pdf: 0x0 pixel, 300dpi, 0.00x0.00 cm, bb=
 \caption{Duality diagram for non-exotic antibaryon-baryon scattering}
 \label{fig:dual8}
\end{figure}

Though invented before the quark model, duality is now better explained with the help of quark diagrams.  Underneath is the Zweig rule, that suppresses the disconnected diagrams. See, e.g., \cite{PhysRevD.64.094507,Roy:2003hk} for an introduction to the Zweig rule, and refs.\ there.  The case of $\bar K N$, or any other non-exotic meson-baryon scattering is shown in Fig.~\ref{fig:dual7}.
For the exotic $KN$ channel the incoming antiquark is $\bar s$, and it cannot annihilate. So there is no possibility of forming a quark-antiquark meson in the $t$ channel, nor a three-quark state in the $s$-channel.  In a famous paper \cite{Rosner:1968si}, Rosner pointed out that as meson-exchanges are permitted in nucleon-antinucleon scattering (or any baryon-antibaryon system with at least one quark matching an antiquark), there should be resonances in the $s$-channel: baryonium was born, and more generally a new family of hadrons.  The corresponding quark diagram is shown in Fig.~\ref{fig:dual8}.
As stressed by Roy \cite{Roy:2003hk}, duality suggests higher exotics.
\subsection{Baryonium in the hadronic-string picture}
\begin{figure}[ht!]
 \centering
 \raisebox{1cm}{\includegraphics[width=.18\textwidth]{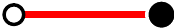}}
 \qquad
 \includegraphics[width=.16\textwidth]{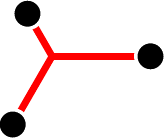}
 \qquad
 \includegraphics[width=.21\textwidth]{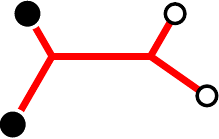}
 \caption{String picture of a meson (left),  a baryons (center) and a baryonium (right)}
 \label{fig:string}
\end{figure}

This concept of duality is illustrated in the hadronic-string picture, which, in turn, is supported by the strong-coupling limit of QCD. See, e.g., the contribution by Rossi and Veneziano in \cite{Montanet:1980te}. A meson is described as a string linking a quark to an antiquark. A baryon contains three strings linking each of the three quarks to a junction, which acts as a sort of fourth component and tags the baryon number. The baryonium has a junction linked to  the two quarks, and another junction linked to the two antiquarks. See Fig.~\ref{fig:string}. The decay happens by string breaking and $q\bar q $, leading either to another baryonium and a meson, or to baryon-antibaryon pair. The decay into two mesons proceeds via the internal annihilation of the two junctions, and is suppressed. 

The baryonium of Jaffe was somewhat similar, with the string realized by the cigar-shape limit of the bag model \cite{Jaffe:1977cv}. Note that the suppression of the decay into mesons is due in this model to a centrifugal barrier, rather than to a topological selection rule.  The orbitally excited mesons consist of a quark and an antiquark  linked by a string, the excited baryons are the analogs with a quark and a diquark, and the baryonia involve a diquark and an antidiquark. 
\subsection{Color chemistry}
Chan Hong-Mo et al.\ \cite{Chan:1978nk} pushed the speculations  a little further in their ``color chemistry''. They have baryonia with color $\bar 3$ diquarks, which decay preferentially into a baryon-antibaryon pair rather than into mesons, also more exotic baryonia in which the diquark has color sextet. Then even the baryon-antibaryon decay is suppressed, and the state is expected to be rather narrow. This was a remarkable occurrence of the color degree of freedom in spectroscopy.  However, there was no indication on how and why such diquark-antidiquark structure arises from the four-body dynamics.
\subsection{Other exotics?}
The baryonium story is  just an episode in the long saga of exotics, which includes the strangeness $S=+1$ ``Z'' baryons in the 60s, their revival under the name ``light pentaquark'' \cite{Tanabashi:2018oca}.  The so-called ``molecular approach''  hadrons was illustrated by the picture of the $\Delta$ resonance as $\pi N$ by Chew and Low \cite{Chew:1955zz}, and of the $\Lambda(1405)$ as  $\bar K N$  by Dalitz and Yan \cite{Dalitz:1960du}, with many further discussions and refinements. 

As reminded, e.g., in \cite{Rossi:2016szw}, there is some analogy between the baryonium of the 70s and 80s and the recent $XYZ$ spectroscopy. The $XYZ$ are mesons with hidden heavy flavor that do not fit in the ordinary quarkonium spectroscopy \cite{Brambilla:2019esw}.  One can replace ``quasi-nuclear'' by ``molecular'', ``baryon number'' by ``heavy flavor'', etc., to translate the concepts introduced for baryonium for use in the discussions about $XYZ$.   The diquark clustering in the light sector is now replaced by an even more delicate assumption, namely $cq$ or $\bar c\bar q$ clustering. While the $X(3872)$ is very well established, some other states either await confirmation or could be  interpreted as mere threshold effects. Before the $XYZ$ wave, it was suggested that baryon-antibaryon states could exist with strange or charmed hyperons. This spectroscopy is regularly revisited. See, e.g., \cite{Maeda:2017ycu} and refs.\ there. 
\section{Antinucleon-nucleon scattering}\label{se:scatt}
In this section, we give a brief survey of  measurements of antinucleon-nucleon scattering and their interpretation, for some final states: $\bar N N$, $\bar \Lambda \Lambda$, and two pseudoscalars. Some emphasis is put on spin observables. It is stressed in other chapters of this book how useful were the measurements done with polarized targets and/or beams for our understanding of the $NN$ interaction, leading to an almost  unambiguous reconstruction of the $NN$ amplitude.
%\footnote{It was a privilege for me to have discussions with pioneers such as F.~Lehar,  J. Bistricki, R. Hess, C. Leluc, \dots}. 
The interest in $\bar NN$ spin observables came at workshops held to prepare the LEAR experiments \cite{Gastaldi:99763,Gastaldi:106464,Amsler:112013}, and at the spin Conference held at Lausanne in 1980 \cite{Joseph:1981zv}. A particular attention was paid to $p\bar p\to\bar \Lambda\Lambda$, but all the theoreticians failed in providing valuable guidance for the last measurements using a polarized target, as discussed below in Sec.~\ref{subsec:scatt:LL}. However,  \textsl{Felix Culpa},%
\footnote{``For God judged it better to bring good out of evil than not to permit any evil to exist'', Augustinus}%
%or \textsl{Virtus Vulnere Virescit},%
%\footnote{``Courage becomes greater through a wound''}
%depending which is your favorite sentence, 
we learned how to better deal with the relationships and constraints among  spin observables. 
\subsection{Integrated cross sections}
As already mentioned, the integrated cross sections have been  measured first at Berkeley, shortly after  the discovery of the antiproton. More data have been taken in many experiments, mainly at the Brookhaven National Laboratory (BNL) and CERN, at various energies.  The high-energy part, together with its proton-proton counter part, probes the Pomerantchuk theorem, Froissart bound and the possible onset of the odderon. See, e.g., \cite{Martynov:2018nyb} and refs.\ there.

\begin{figure}[ht!]
 \centering
 \includegraphics[width=.4\textwidth]{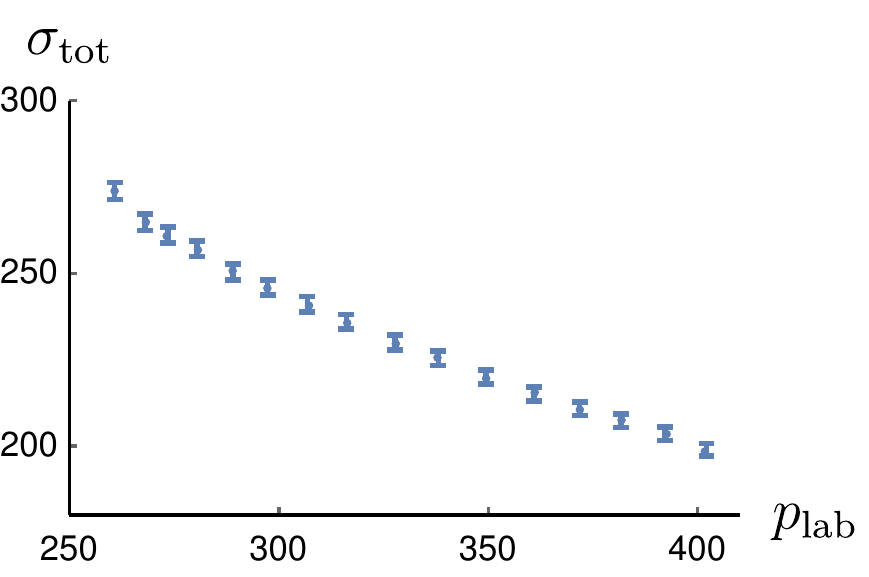}
 % PS172-tot.pdf: 0x0 pixel, 300dpi, 0.00x0.00 cm, bb=
 \caption{Total $\bar p p$ cross section (in mb), as measured by the PS172 collaboration at LEAR.}
 \label{fig:PS172-tot}
\end{figure}

\begin{figure}[ht!]
 \centering
 \includegraphics[width=.4\textwidth]{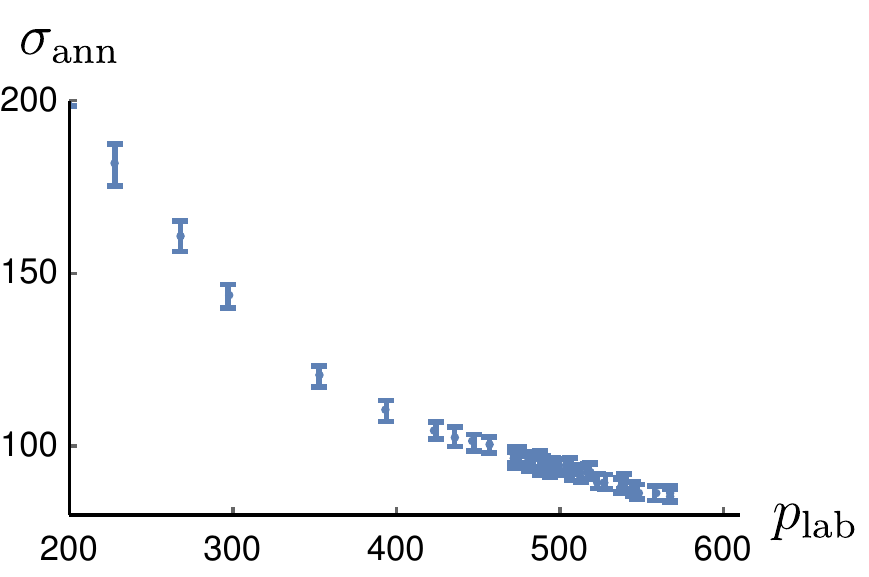}
 % PS173-ann.pdf: 0x0 pixel, 300dpi, 0.00x0.00 cm, bb=
 \caption{Antiproton-proton annihilation cross section (in mb), as measured by th PS173 collaboration at LEAR.}
 \label{fig:PS173-ann}
\end{figure}

As for the low-energy part,  some values  of the total cross section are shown in Fig.~\ref{fig:PS172-tot}, as measured by the PS172 collaboration \cite{Bugg:1987nq}. It can be contrasted to the annihilation cross section of Fig.~\ref{fig:PS173-ann}, due to the PS173 collaboration \cite{Bruckner:1989ew}. When one compares the values at the same energy, one sees that annihilation is more than half the total cross section.  Meanwhile, the integrated charge-exchange cross section is rather small (just a few mb). 

Let us stress once more that the hierarchy $\sigma_\text{ann}>\sigma_\text{el}$ of the annihilation and elastic cross-sections is remarkable. One needs more than a full absorptive core. Somehow, the long-range attraction pulls the wave function towards the inner regions where annihilation takes place \cite{Dalkarov:1977xp,PhysRevC.21.1466}.
\subsection{Angular distribution for elastic and charge-exchange reactions}
\begin{figure}[ht!]
 \centering
 \includegraphics[width=.4\textwidth]{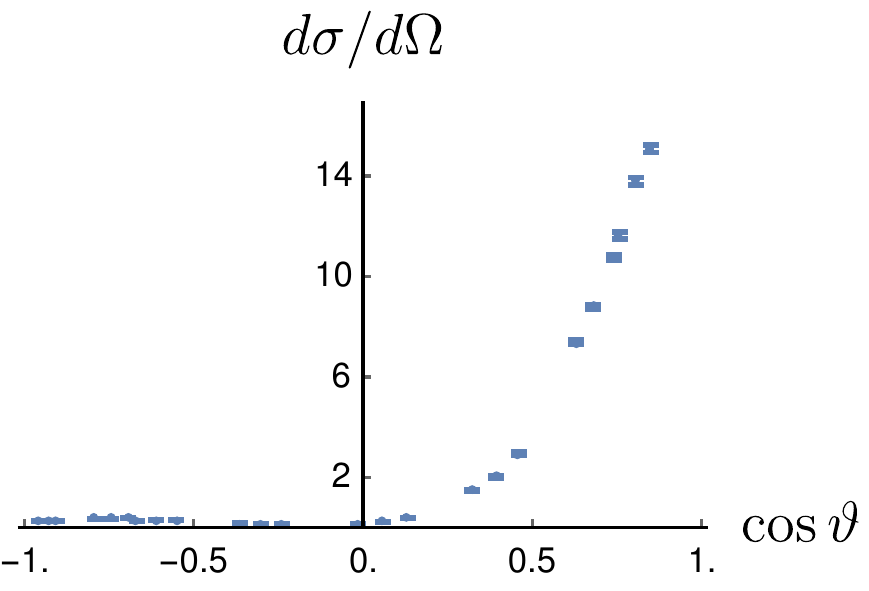}
 % PS198-sig-697.pdf: 0x0 pixel, 300dpi, 0.00x0.00 cm, bb=
 \caption{Angular distribution in elastic $\bar p p\to\bar p p$ scattering at $0.697\,$GeV$/c$, as measured by the PS198 collaboration \cite{Bertini:1989ji}.}
 \label{fig:PS198:sig:697}
\end{figure}

The elastic scattering has been studied in several experiments, most recently at LEAR, in the experiments PS172, PS173, PS198, \dots
An example of differential distribution is shown in Fig.~\ref{fig:PS198:sig:697}.

\begin{figure}[ht!]
 \centering
 \includegraphics[width=.4\textwidth]{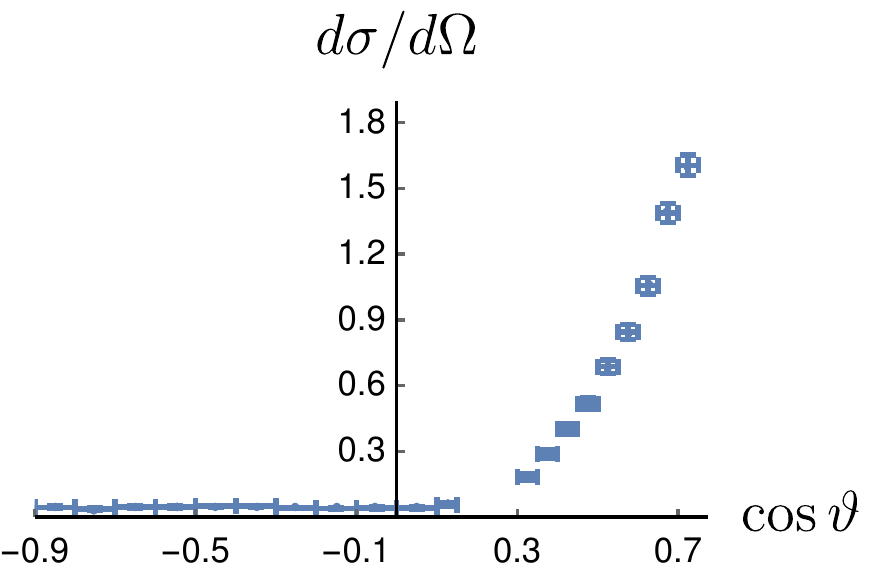}
 % PS199-diff-ce-1083.pdf: 0x0 pixel, 300dpi, 0.00x0.00 cm, bb=
 \caption{Angular distribution for the charge-exchange reaction $\bar  p p\to \bar n n$ at incident momentum $1.083\,$Gev$/c$ in the target frame \cite{Birsa:1994vi}.   Only the statistical error is shown here. Large systematic errors have to be added.}
 \label{fig:PS199-diff}
\end{figure}

The charge exchange scattering has been studied by the PS199-206 collaboration at LEAR. As discussed in one  of the workshops on low-energy antiproton physics \cite{Gastaldi:99763}, charge exchange gives the opportunity to study the interplay between the long-range and short-range physics. An example of differential cross-section is shown in Fig.~\ref{fig:PS199-diff}, published in \cite{Ahmidouch:1995nq}.  Clearly the distribution is far from flat. This illustrates the role of high partial waves. 
\begin{figure}[ht!]
 \centering
 \includegraphics[width=.4\textwidth]{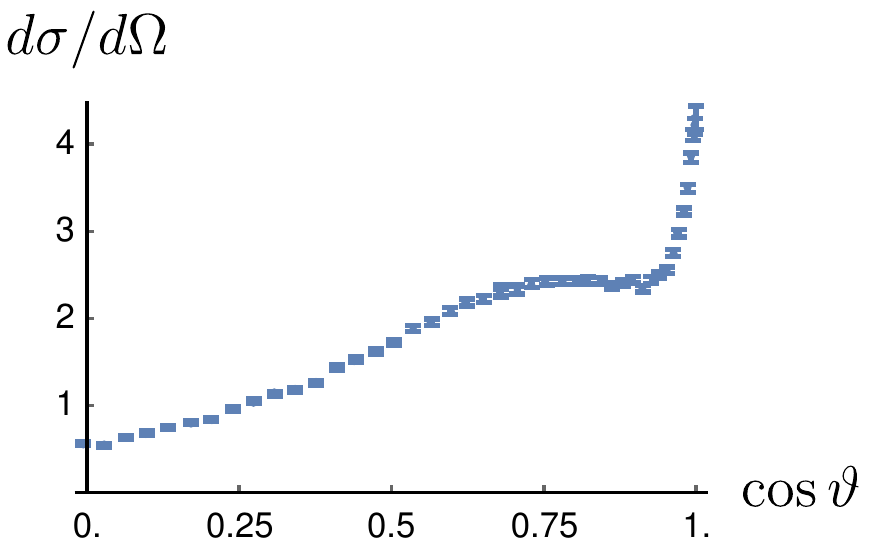}
 % PS199-diff-ce-1083.pdf: 0x0 pixel, 300dpi, 0.00x0.00 cm, bb=
 \caption{Angular distribution for the charge-exchange reaction $\bar  p p\to \bar n n$ at incident momentum $0.601\,$Gev$/c$ in the target frame, as measured by the PS206 collaboration~\cite{Birsa:1994vi}.}
 \label{fig:PS206-diff}
\end{figure}
The amplitude for charge exchange corresponds to the isospin combination
\begin{equation}
\mathcal{M}(\bar p p\to \bar n n)\propto \mathcal{M}_0-\mathcal{M}_1~,
\end{equation}
The smallness of the integrated charge-exchange cross-section is due to a large cancellation in the low-partial waves. But in the high partial waves, there is a coherent superposition. In particular the one-pion exchange gets an isospin factor~$+1$ for $\mathcal{M}_1$, and a factor $-3$ for $\mathcal{M}_0$. 
\subsection{Antineutron scattering}
\begin{figure}[ht!]
 \centering
  \includegraphics[width=.4\textwidth]{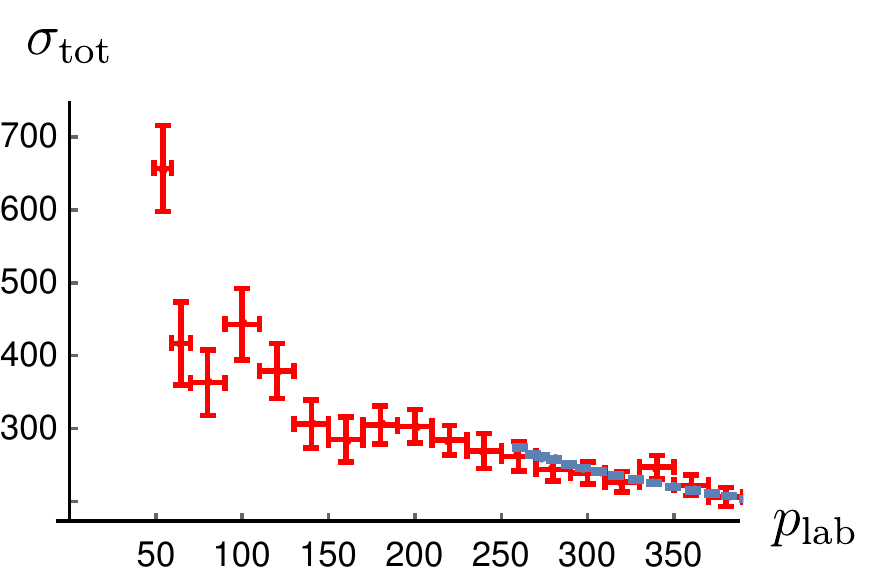}
 % PS201-sig-nbarp.pdf: 0x0 pixel, 300dpi, 0.00x0.00 cm, bb=
 \caption{Total $\bar n p$ cross section (red), as measured by the PS201 collaboration, and comparison with the $\bar p p$ total cross-section (blue)}
 \label{fig:sigbarnp}
\end{figure}

To access to  pure isospin $I=1$ scattering, data have been taken with antiproton beams and deuterium targets, but the subtraction of the $\bar p p$ contribution and accounting for the internal motion and shadowing effects is somewhat delicate.  The OBELIX collaboration at CERN has done direct measurements with antineutrons \cite{Marcello:1999sw}. For instance, the total $\bar n p$ cross-section has been measured between $p_\text{lab}=50$ and 480\,MeV$/c$ \cite{Iazzi:2000rk}. The data are shown in Fig.~\ref{fig:sigbarnp} together with a comparison with the $\bar  p p$ analogs. There is obviously no pronounced isospin dependence.  The same conclusion can be drawn for the $\bar p p$ and $\bar n p$ annihilation cross sections \cite{Klempt:2002ap}. 

\subsection{Spin effects in elastic and charge-exchange scattering}
\begin{figure}[ht!]
 \centerline{
 \includegraphics[width=.45\textwidth]{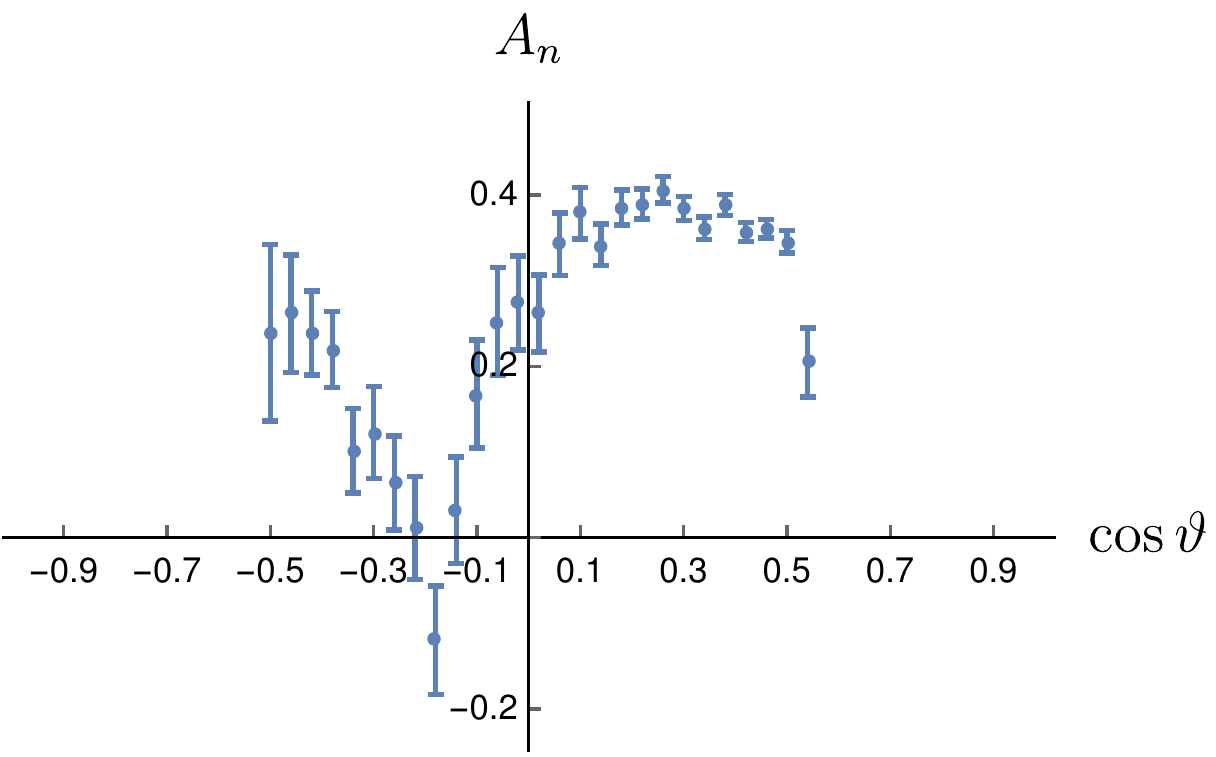}
 \qquad
  \includegraphics[width=.45\textwidth]{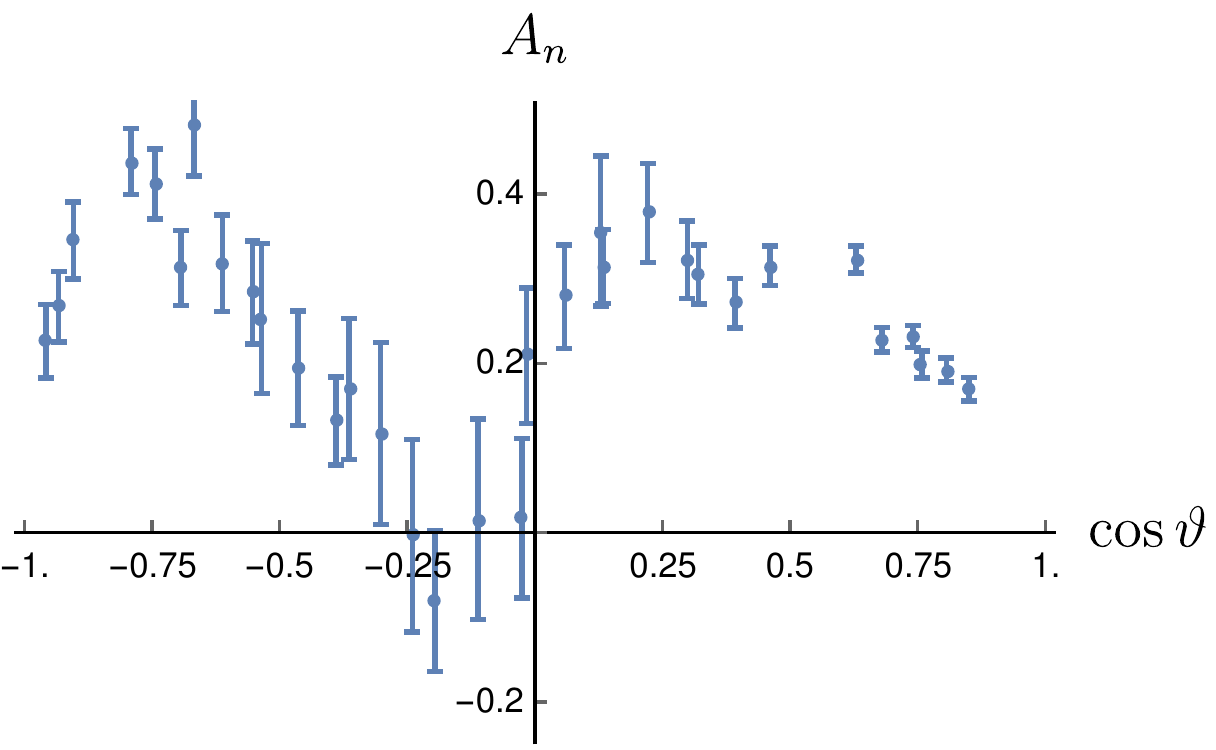}}
 % PS172-679.pdf: 0x0 pixel, 300dpi, 0.00x0.00 cm, bb=
 \caption{Analyzing power of $\bar p p\to \bar p p$ left: at 672\,MeV$/c$, as measured at LEAR by the PS172 collaboration \cite{Kunne:1988tk}, right: at 697\,MeV$/c$ by the PS198 collaboration \cite{Bertini:1989ji}}
 \label{fig:PS172:679}
\end{figure}

\begin{figure}[ht!]
 \centering
 \includegraphics[width=.5\textwidth]{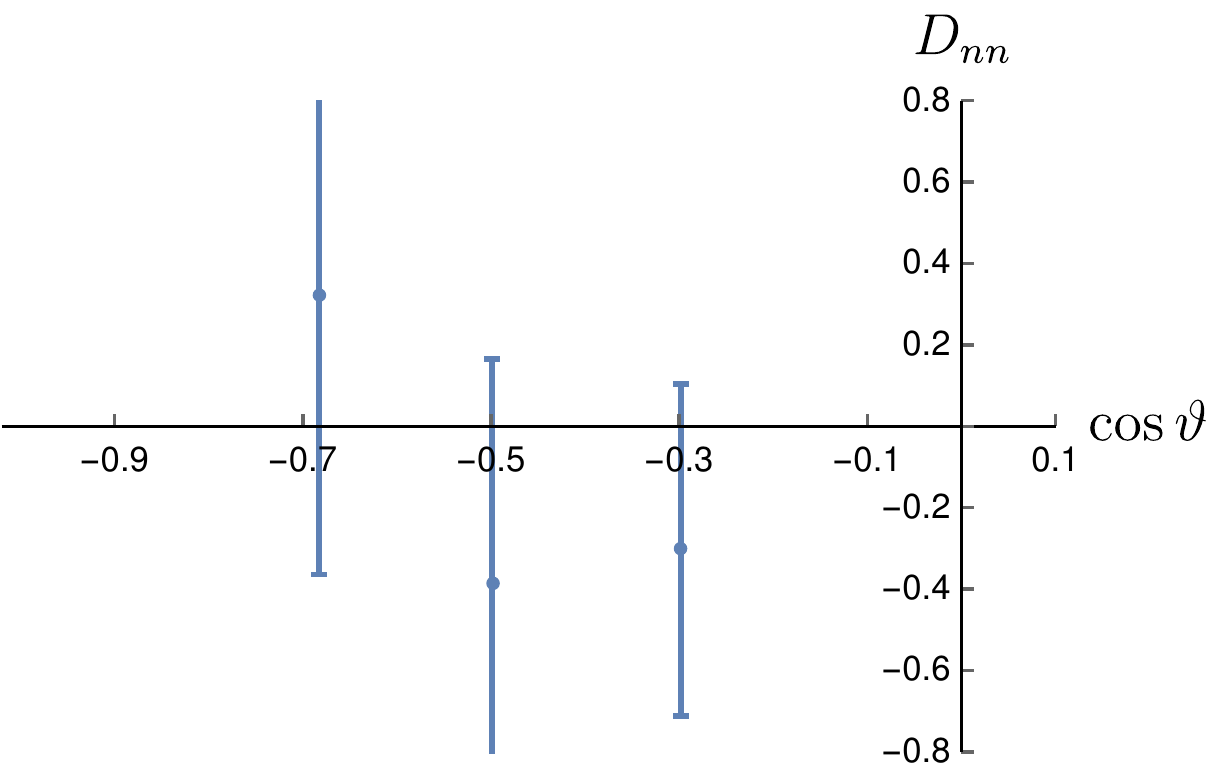}
 % PS172-Dnn-1089.pdf: 0x0 pixel, 300dpi, 0.00x0.00 cm, bb=
 \caption{Transfer of polarization $D_{nn}$ in elastic $\bar p p$ scattering at $p_\text{lab}=1.089\,$GeV$/c$ \cite{Kunne:1991ub}}
 \label{fig:Dnn:el:1089}
\end{figure}

A few measurements of spin effects in $\bar N N\to\bar NN$ were done before  LEAR, mainly dealing with the analyzing power. Some further measurements were done at LEAR, with higher statistics and a wider angular range. 
An example of measurement by PS172 is shown in Fig.~\ref{fig:PS172:679}: the analyzing power of $\bar p p\to \bar p p$ at 679\;MeV$/c$~\cite{Kunne:1988tk}. One can see that the value of $A_n$ is sizable, but not very large. It is compatible with either a moderate spin-orbit component of the interaction, or a rather strong tensor force acting at second order. 
PS172 also measured the depolarization parameter $D_{nn}$ in $\bar p p\to\bar p p$. This parameter $D_{nn}$ expresses the fraction of recoiling-proton polarization along the normal direction that is due to the polarization of the target.   Thus, $D_{nn}=1$ in absence of spin forces. PS172 obtained the interesting result $D_{nn}=-0.169\pm 0.465$
at $\cos\vartheta=-0.169$ for the momentum $p_\text{lab}=0.679\,$GeV$/c$ \cite{Kunne:1991ub}.  The effect persists at higher momentum, as seen in Fig.~\ref{fig:Dnn:el:1089}.

\begin{figure}[ht!]
 \centering
 \includegraphics[width=.5\textwidth]{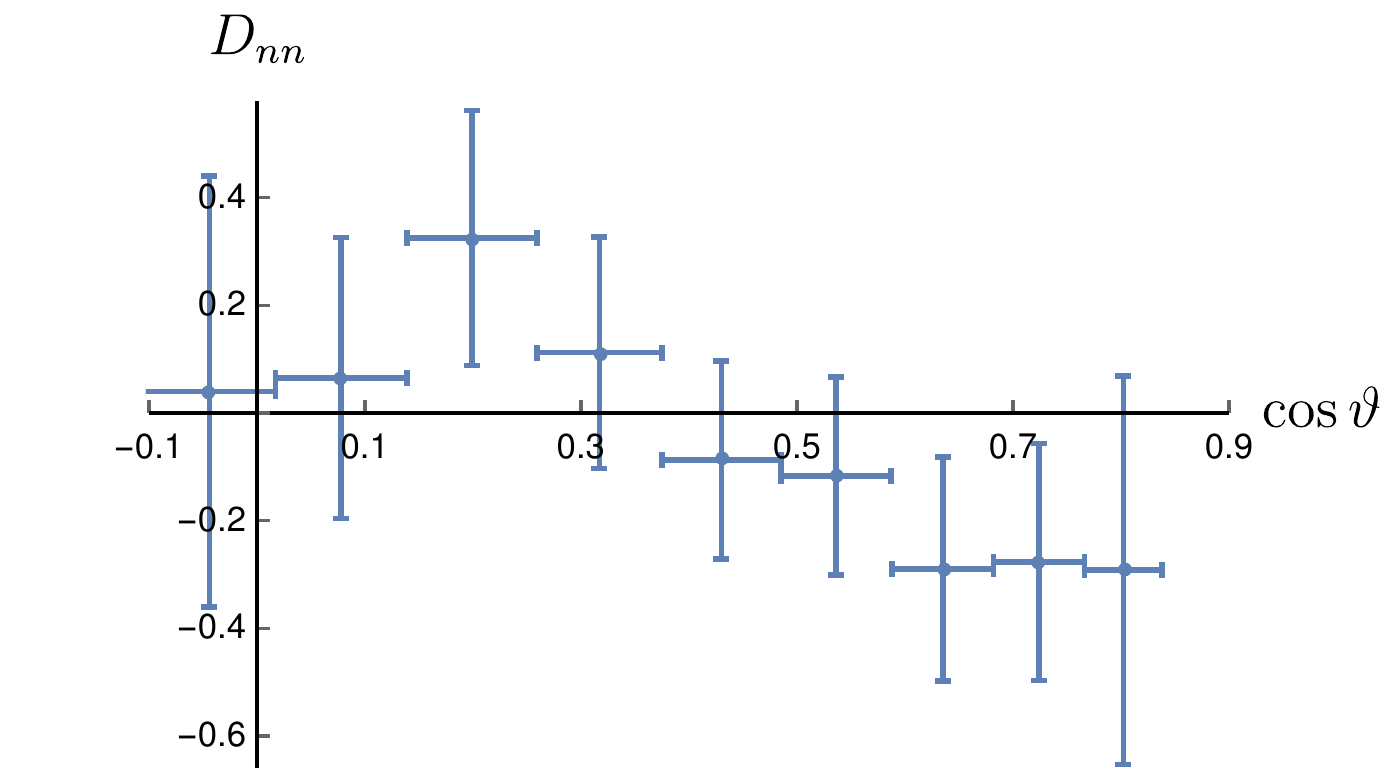}
 % PS199-Dnn-875.pdf: 0x0 pixel, 300dpi, 0.00x0.00 cm, bb=
 \caption{Depolarization parameter in the charge exchange reaction at $0.875\,$GeV$/c$ \cite{Birsa:1993wp}}
 \label{fig:Dnn:ce:875}
\end{figure}

The charge-exchange reaction has been studied by the PS199-206 collaborations at LEAR. See, e.g., \cite{Birsa:1990ta,Birsa:1993wp}. In Fig.~\ref{fig:Dnn:ce:875} is shown the depolarization parameter $D_{nn}$. The effect is clearly  large.  It is predicted that $D_{\ell\ell}$ is even more pronounced, and interestingly, also  $K_{\ell\ell}$,  the transfer of polarization from the target to the antineutron.  This means that one can produce polarized antineutrons by scattering antitprotons on a longitudinally polarized proton target. 
\subsection{Amplitude analysis?}
Decades of efforts have been necessary to achieve a reliable knowledge of the $NN$ interaction at low energy, with experiments involving both a polarized beam and a polarized target. In the case of $\bar N N$, the task is more delicate, as the phase-shifts are complex even at very low energy, and there is no Pauli principle to remove every second partial wave.  So, as we have much less observables available for $\bar N N$ than for $NN$, it is impossible to reconstruct the phase-shifts or the amplitudes: there are unavoidably several solutions with about the same $\chi^2$, and one flips from one solution to another one when one adds or removes a set of data. This is why the fits by Timmermans et al.~\cite{Timmermans:1994pg,Zhou:2012ui} have been received with some skepticism  \cite{Richard:1995xa,Haidenbauer:2019ect}. 

Clearly the measurements of analyzing power and depolarization at LEAR should have been pursued, as was proposed by some collaborations, but unfortunately not approved by the CERN management.  Now, we badly miss the information that would be needed to reconstruct the $\bar NN$ interaction unambiguously, and estimate the possible ways to polarize antiprotons (spin filter, spin transfer). 
\subsection{Potential models}
For the use in studies of the protonium and antinucleon-nucleus systems, it is convenient to summarize the information about the ``elementary'' $\bar N N$ interaction in the form of an effective $\bar N N $ potential. Early attempts were made by Gourdin et al.\ \cite{1958NCim....8..485G}, Bryan and Phillips \cite{Bryan:1968ns} among others, and more recently by Kohno and Weise \cite{Kohno:1986fk}, and the Bonn-J\"ulich group \cite{Haidenbauer:1990cz,Hippchen:1991rr,Mull:1994gz}.  Dover, Richard and Sainio \cite{PhysRevC.21.1466,Dover:1981pp,Richard:1982zr} used as long range potential $V_\text{LR}$ the $G$-parity transformed of the Paris $NN$ potential, regularized in a square-well manner, i.e., $V_\text{LR}(r<r_0)=V_\text{LR}(r_0)$ with $r_0=0.8\,$fm, supplemented by a complex core to account for unknown short-range forces and for annihilation, 
\begin{equation}
 V_\text{SR}(r)=-\frac{V_0+i\,W_0}{1+\exp(-(r-R)/a)}~.
\end{equation}
The short-range interaction was taken as spin and isospin independent,  for simplicity. A good fit of the data was achieved  with two sets of parameters 
\begin{equation}
\begin{aligned}
 \text{model\ DR1}\quad &R=0\phantom{\,\mathrm{fm}}~,\quad &a=0.2\,\mathrm{fm}~,\quad &V_0=21\,\mathrm{GeV}~,\quad  &W_0=20\,\mathrm{GeV}~,\\
 \text{model\ DR2}\quad &R=0.8\,\mathrm{fm}~,\quad &a=0.2\,\mathrm{fm}~,\quad &V_0=0.5\,\mathrm{GeV}~,\quad  &W_0=0.5\,\mathrm{GeV}~.
\end{aligned}
\end{equation}
In \cite{Timmers:1984xv}, the annihilation part is not described by an optical model, but by two effective meson-meson channels.  This probably gives a more realistic energy dependence. 
In some other models , the core contains some spin and isospin-dependent terms, but there are not enough data to constrain the fit.  Some examples are given by the Paris group in \cite{ElBennich:2008vk}, and earlier attempts cited there.  In  \cite{Klempt:2005pp}, a comparison is made of the successive versions of such a  $\bar NN$ potential: the parameters change dramatically when the fit is adjusted to include a new measurement.  The same pattern is observed for the latest iteration \cite{ElBennich:2008vk}. 

More recent models will be mentioned in Sec.~\ref{se:modern} devoted to the modern perspectives, namely an attempt to combine the quark model and meson-exchanges, or potentials derived in the framework of chiral effective theories. 
\subsection{Hyperon-pair production}\label{subsec:scatt:LL}
The PS185 collaboration has measured in detail the reactions of the type $\bar p p\to \bar Y Y'$, where $Y$ or $Y'$ is an hyperon. We shall concentrate here on the $\bar \Lambda\Lambda$ channel, which was commented on by many theorists.  See, e.g., \cite{Kohno:1987uj}.  In the last runs, a polarized hydrogen target was used. Thus $\bar pp\to\bar\Lambda\Lambda$ interaction at low energy is known in great detail, and motivated new studies on the correlations among the spin observables, which are briefly summarized in Appendix \ref{sec:spin:app}.

The weak decay of the $\Lambda$ (and $\bar\Lambda)$ gives access to its polarization in the final state, and thus many results came from the first runs: the polarization $P(\Lambda)$ and $P(\bar\Lambda)$ (which were checked to be equal), and various spin correlations of the final state $C_{ij}$, where $i$ or $j$ denotes transverse, longitudinal, etc.\footnote{The data have been analyzed with the value of the decay parameter $\alpha$ of that time. The parameter $\alpha$ is defined, e.g., in the note ``Baryon decay parameters'' of \protect\cite{Tanabashi:2018oca}. A recent measurement by the BES\RomanNumeralCaps{3} collaboration in Beijing gives a larger value of $\alpha$ \protect\cite{Ablikim:2018zay}. This means that the $\Lambda$ polarization would be  about $17\,\%$ smaller.}\@
In particular the combination
\begin{equation}
 F_0=\frac14\,(1+C_{xx}-C_{nn}+C_{\ell\ell})~,
\end{equation}
corresponds to the percentage of spin singlet, and was found to be compatible with zero within the error bars. Unfortunately, at least two explanations came:
\begin{itemize}
 \item According to the quark model, the spin of $\Lambda$ is carried by the $s$ quark, with the light pair $ud$ being  in  a state spin and isospin zero. 
 The vanishing of the spin singlet fraction is due to the creation of the $s\bar s$ pair in a spin triplet to match the gluon in perturbative QCD or the prescription of the $\SLJ3P0$ model, in which the created quark-antiquark pair has the quantum number $0^{++}$.
 \item In the nuclear-physics type of approach, the reaction is mediated by $K$ and $K^*$ exchanges. This produces a coherence in some spin-triple amplitude, analogous to the strong tensor force in the isospin $I=0$ of $\bar NN$. Hence, the triplet is favored. 
\end{itemize}

It was then proposed to repeat the measurements on a polarized hydrogen target. This suggestion got support and was approved.
In spite of a warning that longitudinal polarization might give larger effect, a transverse polarization was considered as an obvious choice, as it gives access to more observables.  A detailed analysis of the latest PS185 are published in \cite{Bassalleck:2002sd,Paschke:2006za}. 

What retained attention was the somewhat emblematic $D_{nn}$ which measures the transfer of normal polarization from $p$ to $\Lambda$ (in absence of spin effects, $D_{nn}=1$). It was claimed that the transfer observable $D_{nn}$ could distinguish among the different scenarios for the dynamics \cite{Haidenbauer:1992gv}, with quark models favoring  $D_{nn}$ positive (except models making use of a polarized $s\bar s$ sea \cite{Alberg:1995zp}), and meson-exchange $D_{nn}<0$.  
When the result came with $D_{nn}\sim 0$,  this was somewhat a disappointment.  But in fact, it was realized \cite{Richard:1996bb,2009PhR...470....1A} that $D_{nn}\sim 0$ was a \emph{consequence} of the earlier data! As reminded briefly in appendix \ref{sec:spin:app}, there are indeed many constraints among the various spin observables of a given reaction. 
For instance, one can show that 
\begin{equation}
 C_{\ell\ell}^2+D_{nn}^2\le 1~.
\end{equation}
This inequality, and other similar constraints, implied that $D_{nn}$ had be small, just from data taken with an unpolarized target, while $D_{\ell\ell}$ had a wider permitted range. 

\begin{figure}[ht!]
 \centerline{
 \includegraphics[width=.49\textwidth]{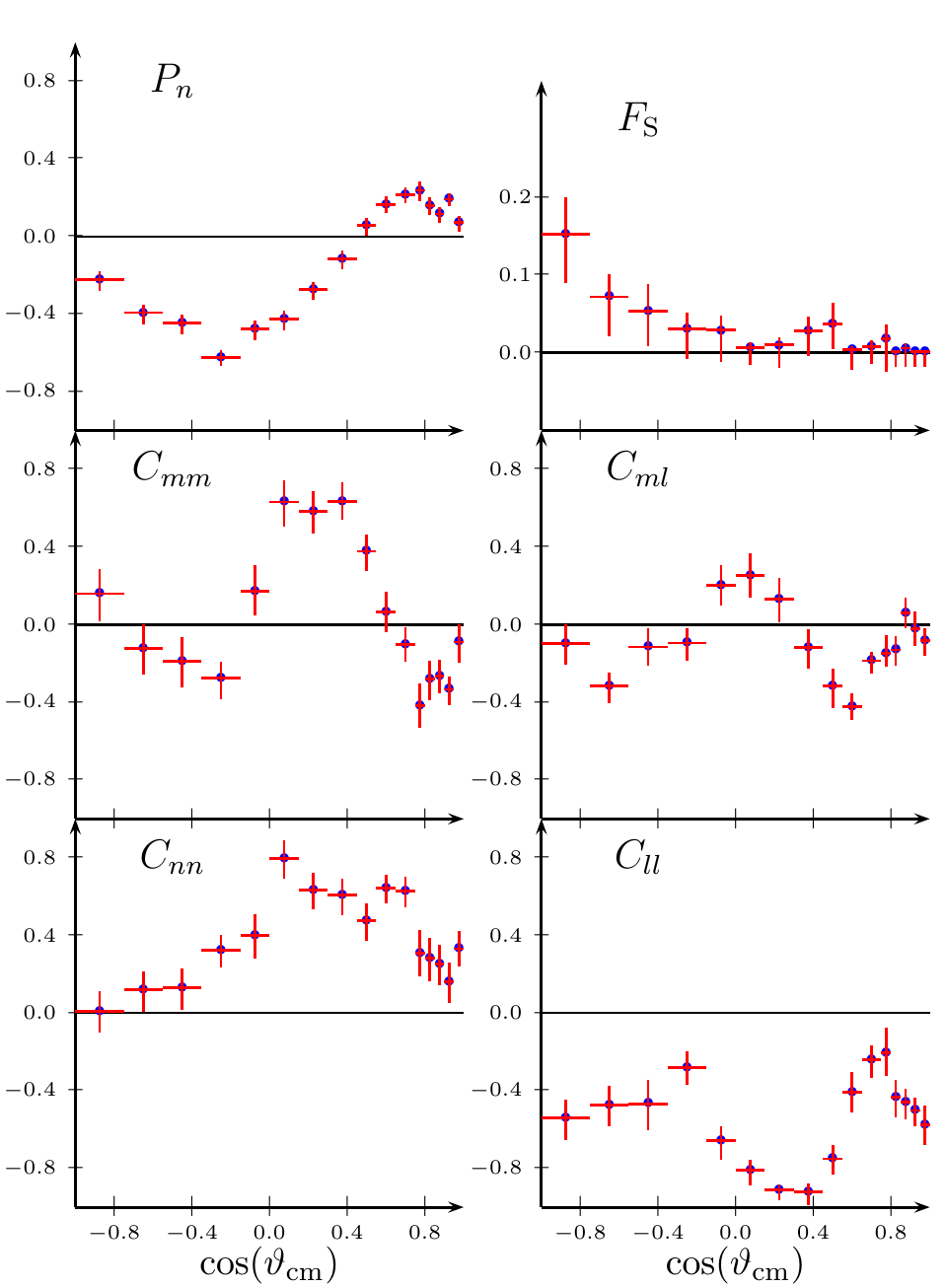}
 \hfill
 \includegraphics[width=.49\textwidth]{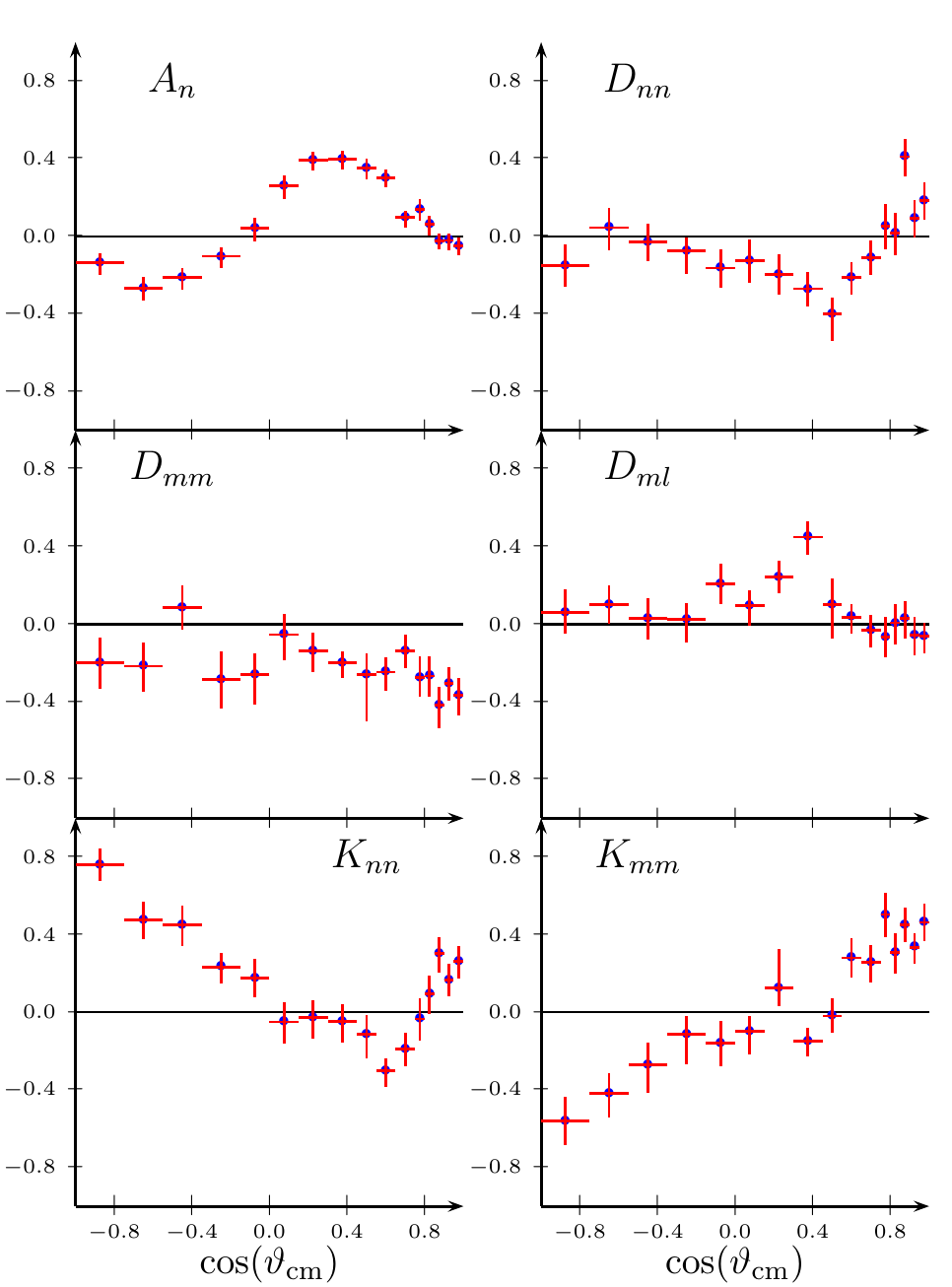}}
 % PS185a.pdf: 0x0 pixel, 300dpi, 0.00x0.00 cm, bb=
 \caption{\label{fig:PS185}Some spin observables of the reaction $\bar p p\to \bar\Lambda\Lambda$}
\end{figure}

A sample of the PS185 results can be found in Fig.~\ref{fig:PS185}.
\subsection{Spin effects in annihilation into two pseudoscalar mesons}\label{subse:ann:2ps}
The reactions $\bar p p\to \pi^+\pi^-$ (and to a lesser extent $\pi^0\pi^0$) and $K^+K^-$ were measured before LEAR. For instance, some results can be read in the proceedings of the Strasbourg conference in 1978 \cite{Fridman:101482}.  However, some adventurous analyses concluded to the existence of unnatural-parity broad resonances, the large-width sector of baryonium. Needless to say that such analyses with few or no spin observables, were flawed from the very beginning. The same methods, and sometimes the same authors, were responsible for the misleading indication in favor of the so-called $Z$ baryons with strangeness $S=+1$, the ancestor of the late light pentaquark $\theta(1540)$. 

\begin{figure}[ht!]
 \centering
 \includegraphics[width=.7\textwidth]{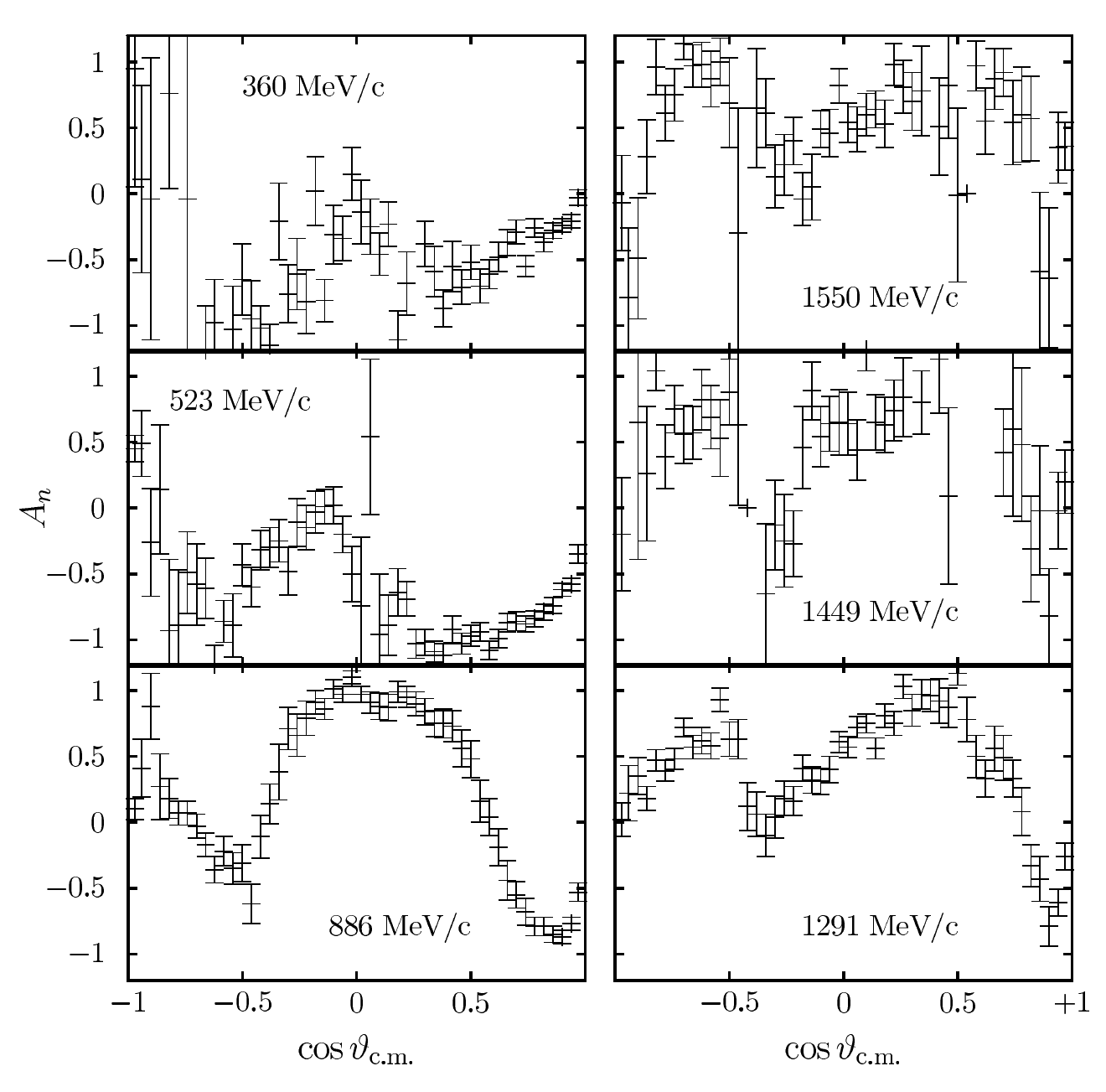}
 % Polpipi.pdf: 0x0 pixel, 300dpi, 0.00x0.00 cm, bb=
 \caption{Some results on $\bar p p\to \pi\pi$ polarization at LEAR}
 \label{fig:polpipi}
\end{figure}

\begin{figure}[ht!]
 \centering
 \includegraphics[width=.7\textwidth]{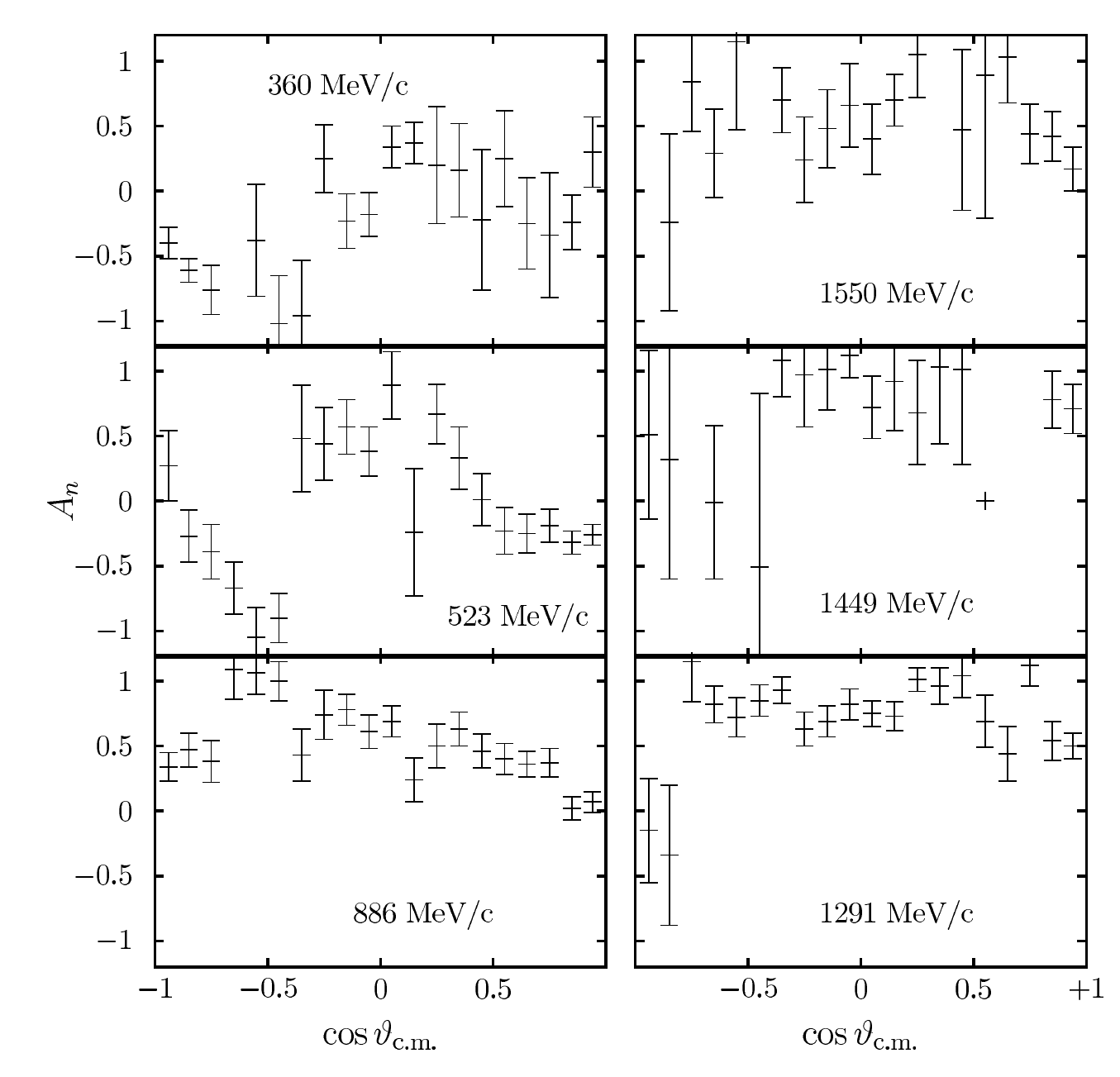}
 % Polpipi.pdf: 0x0 pixel, 300dpi, 0.00x0.00 cm, bb=
 \caption{Some results on $\bar p p\to \bar K K$ polarization at LEAR}
 \label{fig:polKK}
\end{figure}

The LEAR experiment PS172 remeasured these reactions with a polarized target.  This gives access to the analyzing power $A_n$ , the analog of the polarization in the crossed reactions such as $\pi^- p\to\pi^-p$.   Remarkably, $A_n$ is very large, in some wide ranges of energy and angle.  See Figs.~\ref{fig:polpipi} and \ref{fig:polKK}. There is a choice of amplitudes, actually the transversity amplitudes, such that
\begin{equation}\label{eq:An}
 A_n=\frac{|f]^2-|g|^2}{|f]^2+|g|^2}~,
\end{equation}
In this notation,  $|A_n|\sim 1$ requires one amplitude $f$ or $g$ to be dominant.  This was understood from the coupled channel effects 
\cite{Liu:1990rd,Takeuchi:1992si}. Alternatively, one can argue that the initial state is made of partial waves 
$\SLJ{3}{(J-1)}{J}$ and $\SLJ{3}{(J+1)}{J}$ coupled by tensor forces. The amplitudes $f$ and $g$ correspond to the eigenstates of the tensor operator $S_{12}$ (see Sec. \ref{se:NN-to-NbarN}), and the amplitude in which the tensor operator is strongly attractive tends to become dominant \cite{Elchikh:1993sn}

\begin{subappendices}
 \subsection{Appendix: Constraints on spin observables}\label{sec:spin:app}
\begin{figure}[ht!]
 \centerline{%
 \includegraphics[width=.35\textwidth]{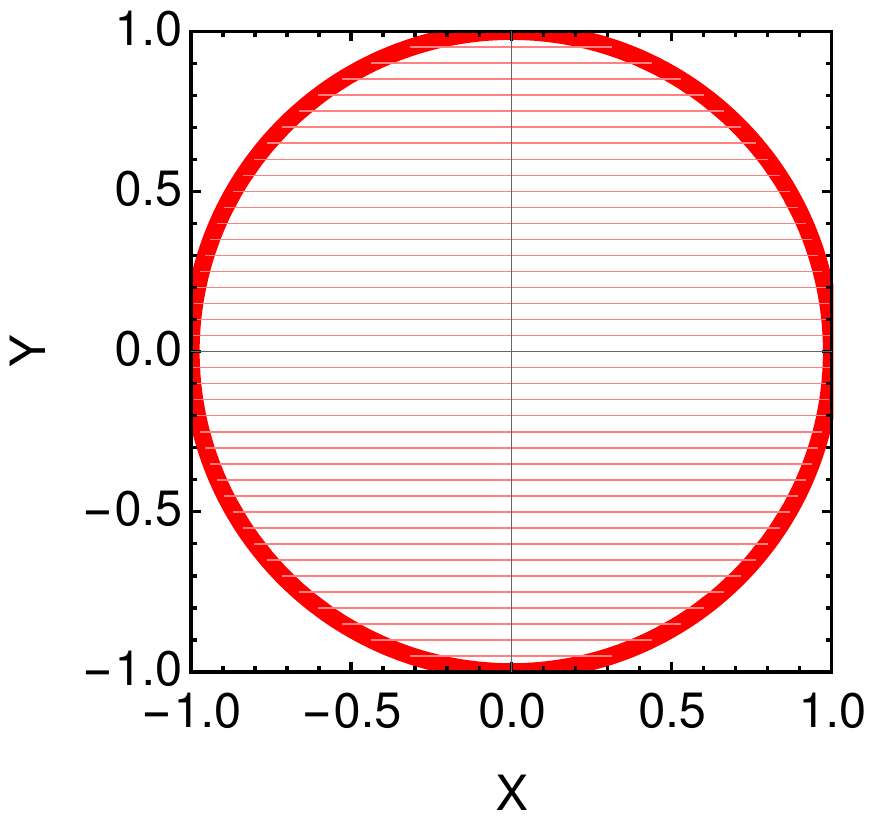}
 \qquad
 \includegraphics[width=.35\textwidth]{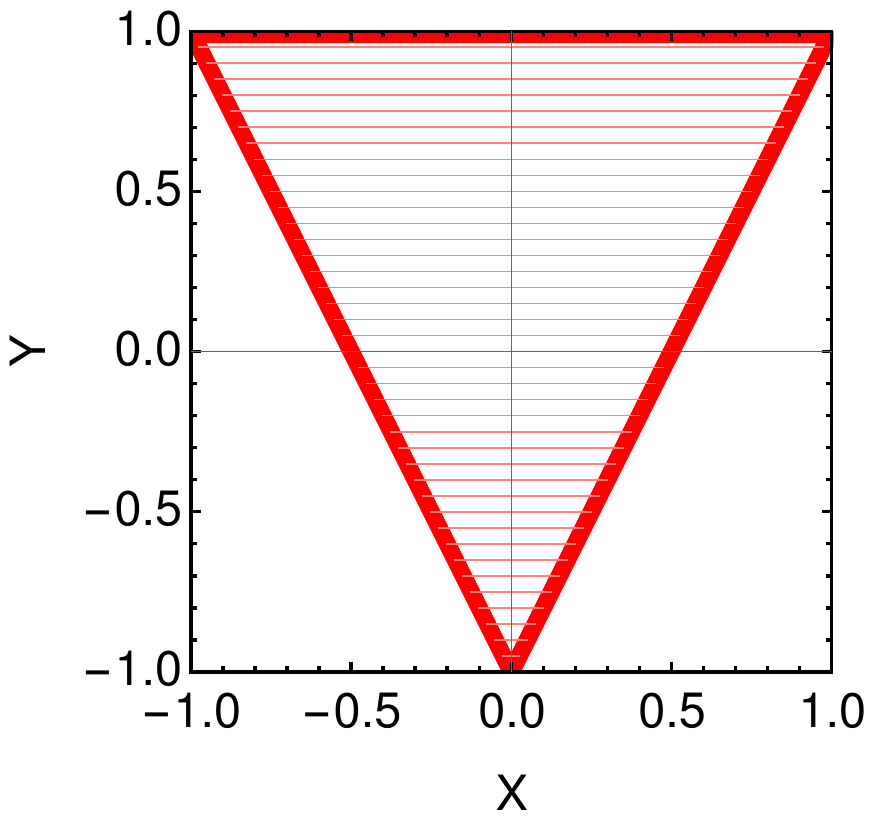}}
 % spindomain1.pdf: 0x0 pixel, 300dpi, 0.00x0.00 cm, bb=
 \caption{Examples of constraints among spin observables: only the colored area is permitted}
 \label{fig:spin-domain}
\end{figure}

A typical spin observable $X$ is usually normalized such that $-1\le X \le +1$. But if one considers two normalized observables $X$ and $Y$ of the same reaction, several scenarios can occur:
\begin{itemize}
 \item The entire square $-1\le X,\,Y \le +1$ is allowed. Then the knowledge of $X$ does not constrain $Y$. 
 \item $\{X,Y\}$ is restricted to a subdomain of the square. One often encounters the unit circle $X^2+Y^2\le 1$. In such case a large $X$ implies a vanishing $Y$.
 This is what happens for $D_{nn}$ vs.\ some of $C_{ij}$ in $\bar p p\to \bar\Lambda\Lambda$.  Another possibility is a triangle, see Fig.~\ref{fig:spin-domain}.
\end{itemize}

For instance, in the simplest case of $\pi N\to \pi N$ (or its cross reaction as in Eq.~\ref{eq:An}), there is a set of amplitudes such that the polarization (or the analyzing power), and the two independent transfer of polarization) are given by 
\begin{equation}
 X=\frac{|f]^2-|g|^2}{|f]^2+|g|^2}~,\quad Y=\frac{2\ \RE(f^*\,g)}{|f]^2+|g|^2}~,\quad
 Z=\frac{2\,\IM(f^*\,g)}{|f]^2+|g|^2}~,
\end{equation}
such that $X^2+Y^2+Z^2=1$ and thus $X^2+Y^2\le 1$. For reactions with two spin-1/2 particles, the algebra is somewhat more intricate  \cite{2009PhR...470....1A}. 

At about the same time as the analysis of the PS172 and PS185 data, similar inequalities were derived for the spin-dependent parton distributions, in particular by the late Jacques Soffer, starting from the requirement of positivity. An unified presentation of the inequalities in the hadron-hadron and quark distribution sectors can be found in \cite{2009PhR...470....1A}. The  domain allowed for three normalized observables $X$, $Y$, $Z$ can be found in this reference, with sometimes rather amazing shapes for the frontier. 

Perhaps a new strategy could emerge. Instead  of either disregarding all spin measurements, or to cumulate all possible spin measurements in view of an elusive full reconstruction, one could advocate a stage by stage approach: measure first a few observables and look for which of the remaining are less constrained, i.e., keep the largest potential of non-redundant information.
\end{subappendices}
\section{Protonium}\label{se:protonium}
Exotic atoms provide a subtle investigation of the hadron-nucleon and hadron-nucleus interaction at zero energy.  For a comprehensive review, see \cite{Deloff:2003ns}. Let us consider $(h^-,A)$, where $h^-$ is a negatively charged hadron such as $\pi^-$ or $K^-$, and $A$ a nucleus of charge $+Z$. One can calculate the energy levels $E_{n,\ell}^{(0)}$ by standard QED techniques, including finite volume, vacuum polarization, etc.  The levels are shifted and broadened  by  the strong interactions, and it can be shown (most simply in potential models, but also in effective theories), that the complex shift is given by
\begin{equation}\label{eq: Deser-T}
 \delta E_{n,\ell}=E_{n,\ell}-E_{n,\ell}^{(0)}\simeq C_{n,\ell}\, a_\ell~,
\end{equation}
 where $a_{\ell}$ is the scattering length for $\ell=0$, volume for $\ell=1$, \dots of the strong $h\,A$ interaction. $C_{n,\ell}$ is a know constant involving the reduced mass and the $\ell^\text{th}$ derivative of the radial  wave function at the origin of the pure Coulomb problem.  
 Experiments on protonium have been carried out before and after LEAR. For a summary, see, e.g., \cite{Klempt:2002ap}. The latest results are:
 \begin{itemize}
  \item For the 1S level, the average shift and width are \cite{AUGSBURGER1999149} $\delta E(1\mathrm{S})=712.5\pm20.3\,$eV (to be compared with the Bohr energy $E(1\mathrm{S})\simeq -12.5\,$keV), and $\Gamma(1\mathrm{S})=1054\pm 65\,$eV,
   with a tentative separation of the hyperfine level as 
 $\delta E(\SLJ3S1)=785\pm35\,$eV,  $\Gamma(\SLJ3S1)=940\pm80\,$eV and 
  $\delta E(\SLJ1S0)=440\pm75\,$eV,  $\Gamma(\SLJ1S0)=1200\pm250\,$eV.
  The repulsive charcater is a consequence of the strong annihilation.
  \item For the 2P level, one can not distinguish among $\SLJ1P1$, $SLJ3P1$ and $\SLJ3P2$, but this set of levels is clearly separated from the $\SLJ3P0$ which receives a larger attractive shift, as predicted in potential models (see, e.g., \cite{Richard:1982zr,Entem:2006dt} and a larger width. More precisely
  \cite{GOTTA1999283}, $\delta E [2(\SLJ3P2,\SLJ31P1,\SLJ3P1)]\simeq 0$, $\Gamma [2(\SLJ3P2,\SLJ31P1,\SLJ3P1)]=38\pm9\,$meV, and  $\delta E [2\SLJ3P0]\simeq -139\pm28\,$mEV, $\Gamma [2\SLJ3P0]=489\pm 30\,$meV. For the latter, the admixture of the $\bar nn$ component is crucial in the calculation, and the wave function at short distances is dominated by it isospin $I=0$ component \cite{Dover:1991mu}. 
 \end{itemize}
\subsection{Quantum mechanics of exotic atoms}
Perturbation theory is valid if the energy shift is small as compared to the level spacing. However, a small shift does not mean that perturbation theory is applicable.
 For instance, a hard core of radius $a$ added to the Coulomb interaction gives a small upward shift to the  levels, as long as the core radius $a$ remains small as compared to the Bohr radius $R$, but a naive application of ordinary perturbation theory will give an infinite correction! For a long-range interaction modified by a strong short-range term, the expansion parameters is the ratio of the ranges, instead of the coupling constant.  At leading order, 
the energy shift is given by the formula of Deser et al. \cite{Deser:1954vq}, and Trueman \cite{Trueman:1961zza}, which reads
 \begin{equation}\label{eq:D-T}
  \delta E \simeq 4\,\pi\,|\phi_{n\ell}(0)|^2\,a_0~,
 \end{equation}
where $a_0$ is the scattering length in the short-range potential alone, and $\phi_{n\ell}(0)$ the unperturbed wave function at zero separation. For a simple proof, see, e.g., Klempt et al.~\cite{Klempt:2002ap}.
The formula \eqref{eq:D-T} and its generalization \eqref{eq: Deser-T} look perturbative, because of the occurrence of the unperturbed wavefunction, but it is not, as the scattering length (volume, \dots) $a_\ell$ implies iterations of the short-range potential. 

 There are several improvements and generalizations  to any superposition of a short-range and a long-range potential, the latter not necessarily Coulombic, see, e.g.,~ \cite{Combescure:2007ki}. For instance, in the physics of cold atoms, one often considers systems experiencing some harmonic confinement and a short-range pairwise interaction. 
\subsection{Level rearrangement}
\begin{figure}[ht!]
 \centering
 \includegraphics[width=.5\textwidth]{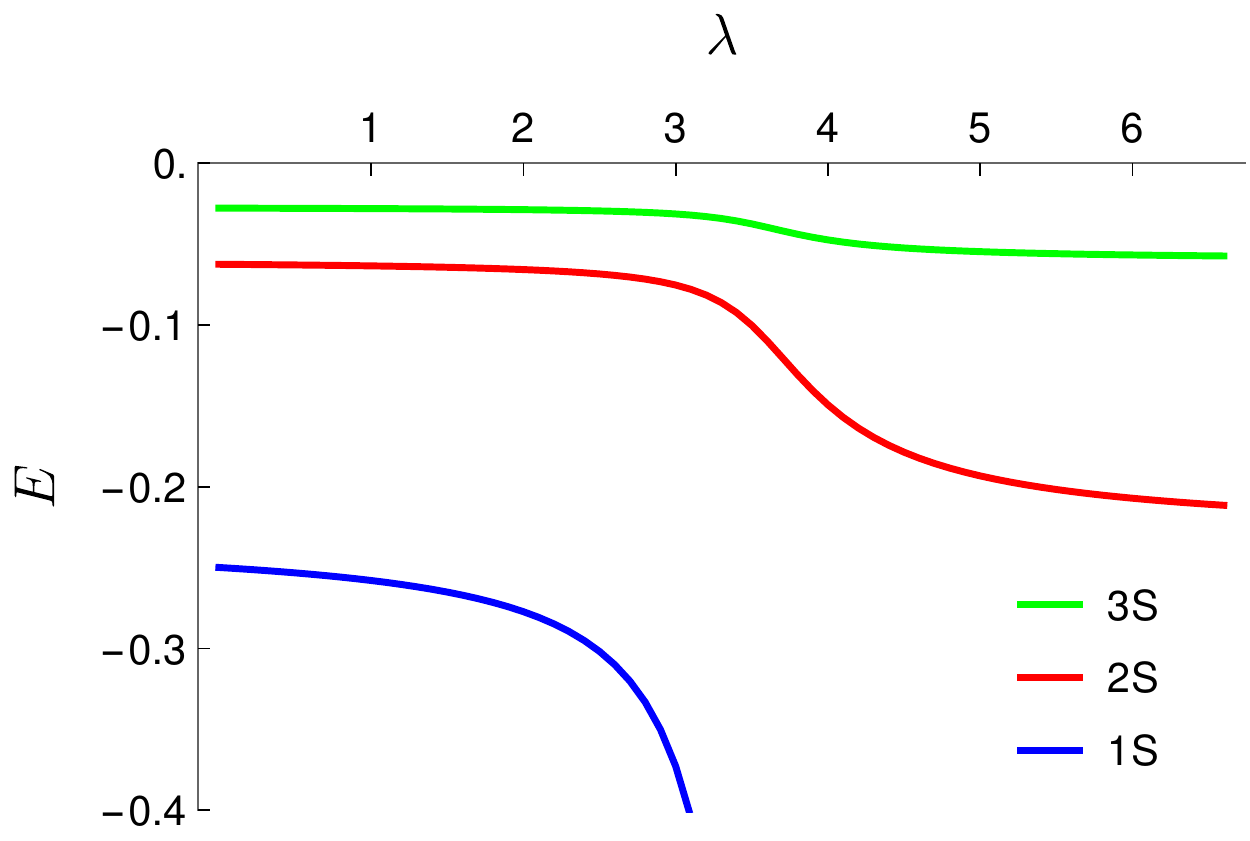}
 % Rearr-100.pdf: 0x0 pixel, 300dpi, 0.00x0.00 cm, bb=
 \caption{Rerrangement of levels: first three levels of the radial equation $-u''(r)+V(r)\,u(r)=E\,u(r)$ with $V(r)=-1/r+\lambda\,a^2 \exp(-a\,r)$, here with $a=100$ and $\lambda$ variable.}
 \label{fig:proto}
\end{figure}

The approximation \eqref{eq:D-T} implies that the scattering length $a$ remains small as compared to the Bohr radius (or, say, the typical size of the unperturbed wave function). Zel'do\-vich~\cite{Zeldo:1960}, Shapiro \cite{Shapiro:1978wi} and others have studied what happens, when the attractive short-range potential becomes large enough to support a bound state on its own. Let the short-range attractive interaction be $\lambda\,V_\text{SR}$, with $\lambda>0$.  When $\lambda$ approaches and passes the critical value $\lambda_0$ for the first occurrence of binding in this potential, the whole Coulomb spectrum moves rapidly.  The 1S state drops from the keV to the MeV range, the 2S level decreases rapidly and stabilizes in the region of the former 1S, etc. 
See, for instance, Fig.~\ref{fig:proto}. Other examples are given in \cite{Deloff:2003ns,Combescure:2007ki}.

 It was then suggested that a weakly bound quasi-nuclear $\bar N N$ state will be revealed by large shifts in the atomic spectrum of protonium \cite{Shapiro:1978wi}.  However, this rearrangement scenario  holds  for a single-channel real potential $V_\text{SR}$.  In practice, the potential is complex, and the Coulomb spectrum is in the $\bar p p$ channel, and the putative baryonium in a state of pure isospin $I=0$ or $I=1$. Hence, the rearrangement pattern is more intricate. 
\subsection{Isospin mixing}
In many experiments dealing with ``annihilation at rest'', protonium is the initial state before the transition $N\bar N\to\,$mesons.  Hence the phenomenological analysis include parameters describing the protonium: S-wave vs.\ P-wave probability and  isospin mixing.  Consider, e.g., protonium in the ${}^1\mathrm{S}_0$ state. In a potential model, its dynamics is given by
\begin{equation}
\begin{aligned}
 -u''(r)/m+ V_{11}\, u(r)+V_{12}\, v(r) -\frac{e^2}{r}\,u(r)&=E_{1,0}\,u(r)~,\\
 -v''(r)/m+ V_{22}\, v(r)+V_{21}\, u(r) +2\,\delta m\,u(r)&=E_{1,0}\,v(r)~,
\end{aligned}  
\end{equation}
where $\delta m$ is the mass difference between the proton and the neutron, and the strong (complex) potentials are the isospin combinations
\begin{equation}
V_{11}=V_{22}=\frac12(V_{I=0}+V_{I=1})~,\qquad
V_{12}=V_{21}=\frac12(V_{I=0}-V_{I=1})~.
\end{equation}
The energy shift is well approximated by neglecting the neutron-antineutron component, i.e., $v(r)=0$. But at short distance, this component is crucial. In most current models, one isospin component is dominant, so that the protonium wave function is dominantly either $I=0$ or $I=1$ at short distances, where annihilation takes place. This influences much the pattern of branching ratios.  For instance, Dover et al. \cite{Dover:1991mu} found in a typical potential model that the $\SLJ3P0$ level consists of $95\,\%$ of isospin $I=0$ in the annihilation region. For $\SLJ3P1$, the $I=1$ dominates, with  $87\,\%$.
See  \cite{Dover:1991mu,Carbonell:1992wd,Furui:1990ex} for a detailed study of the role  of the $\bar n n$ channel on the protonium levels and their annihilation.  

\subsection{Day-Snow-Sucher effect}\label{subse:DSN}
When a low-energy antiproton is sent on a gaseous or liquid hydrogen target, it is further slowed down by electromagnetic interaction, and is captured in a high orbit of the antiproton-proton system.  The electrons are usually expelled during the capture and the subsequent decay of the antiproton toward lower orbits. The sequence favors circular orbits with $\ell=n-1$, in the usual notation.  Annihilation is negligible  for the high orbits,  and becomes about $1\%$ in 2P and, of course, $100\%$ in  $1S$.  This was already predicted in the classic paper by Kaufmann and Pilkuhn \cite{Kaufmann:1978vp}.

In a dense target, however, the compact $\bar p p$ atom travels inside the orbits of the ordinary atoms constituting the target, and experiences there an electric field which,  by Stark effect, mixes the $(\ell=n-1, n)$ level with states of same principal quantum number $n$ and lower orbital momentum.  Annihilation occurs from the states with the lowest $\ell$. This is known as the  Day-Snow-Sucher effect \cite{Day:1960zz}.  In practice, to extract the branching ratios and distinguish $S$-wave from $P$-wave annihilation, one studies the rates as a function of the target density.
\subsection{Protonium ion and protonium molecule }
So far, the physics of hadronic atoms has been restricted to 2-body systems such as $\bar p p$ or $K^- A$. In fact, if one forgets about the experimental feasibility, there are many other possibilities. If one takes only the long-range Coulomb interaction, without electromagnetic annihilation nor strong interaction, many stable configurations exist, such as $\bar p p p$, the protonium ion, or $\bar p\bar p p p$, the heavy analog of the positronium molecule. Identifying these states and measuring the shift and witdh of the lowest level would be most interesting. Today this looks as science fiction, as it was the case when $\mathrm{Ps}_2=e^+e^+e^-e^-$ was suggested by Wheeler in 1945. But $\mathrm{Ps}_2$ was eventually detected, in 2007.
\section{The antinucleon-nucleus interaction}\label{se:Nbar-A}
\subsection{Antinucleon-nucleus elastic scattering}\label{subse:pbar-A}
\begin{figure}[ht!]
 \centerline{
\includegraphics[width=.3\textwidth]{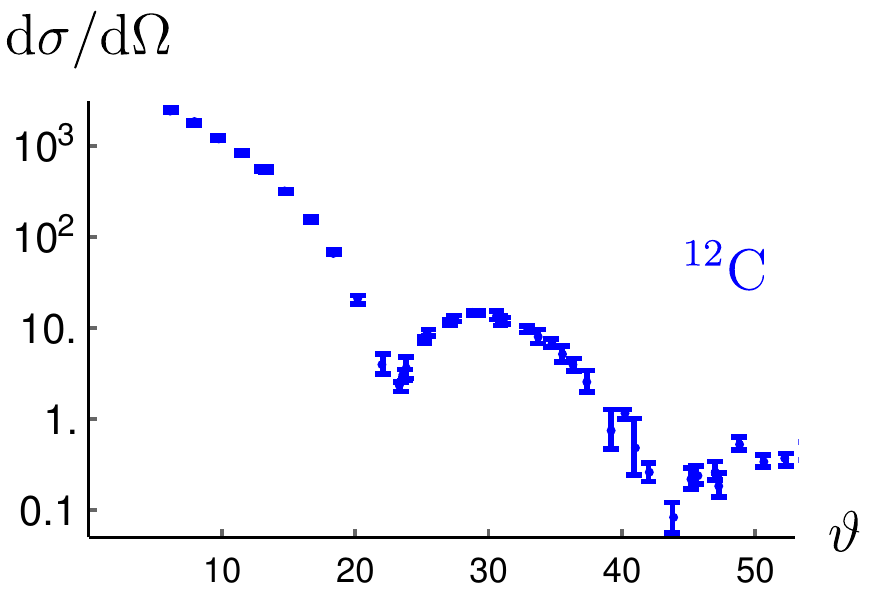}
\qquad \includegraphics[width=.3\textwidth]{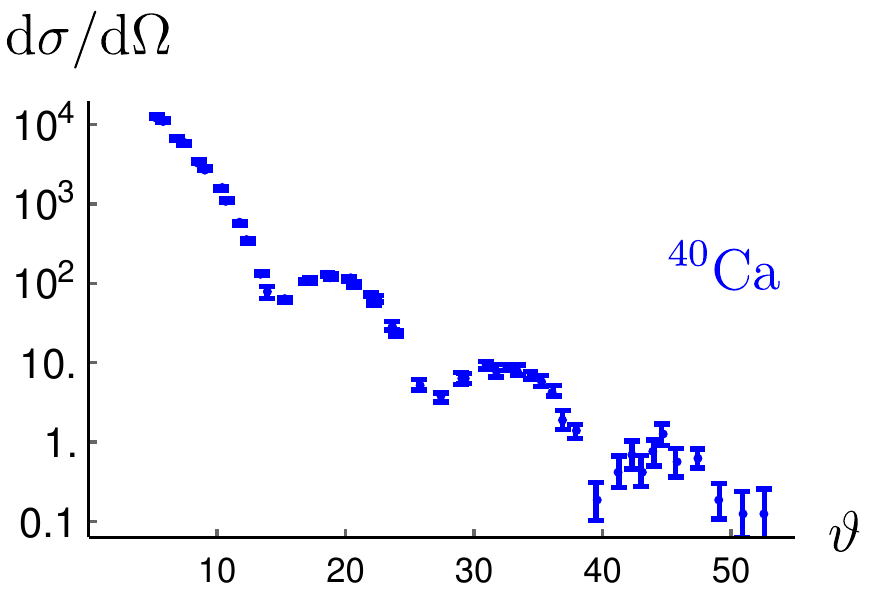}\qquad
\includegraphics[width=.3\textwidth]{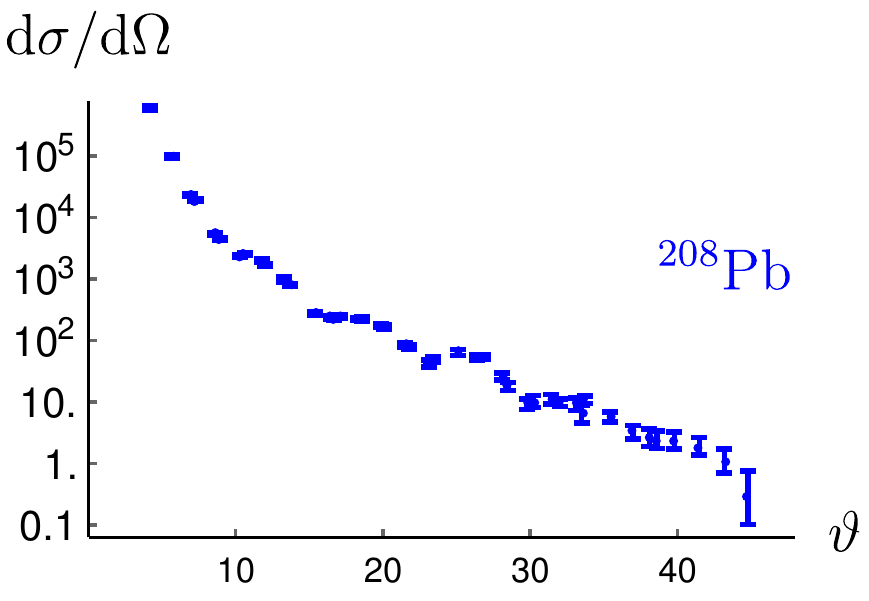}}
% antip-12C-179.pdf: 0x0 pixel, 300dpi, 0.00x0.00 cm, bb=
 \caption{Angular distribution for $\bar p$ scattering on the nuclei \isotope[12]{C}, \isotope[40]{Ca} and \isotope[208]{Pb} at kinetic energy $T_{\bar p}=180\,$MeV \cite{Garreta:1984rs}.}
 \label{fig:PS184}
\end{figure}

At the very beginning of LEAR, Garreta et al.~\cite{Garreta:1984rs} measured the angular distribution of $\bar p$-$A$ scattering, where $A$ was \isotope[12]{C},  \isotope[40]{Ca} or \isotope[208]{Pb}. Some of their  results are reproduced in Fig.~\ref{fig:PS184}.% 
\footnote{I thank Matteo Vorabbi for making available his retrieving of the data in a convenient electronic form}\@
More energies and targets were later measured.
\begin{figure}[ht!]
 \centering
 \includegraphics[width=.4\textwidth]{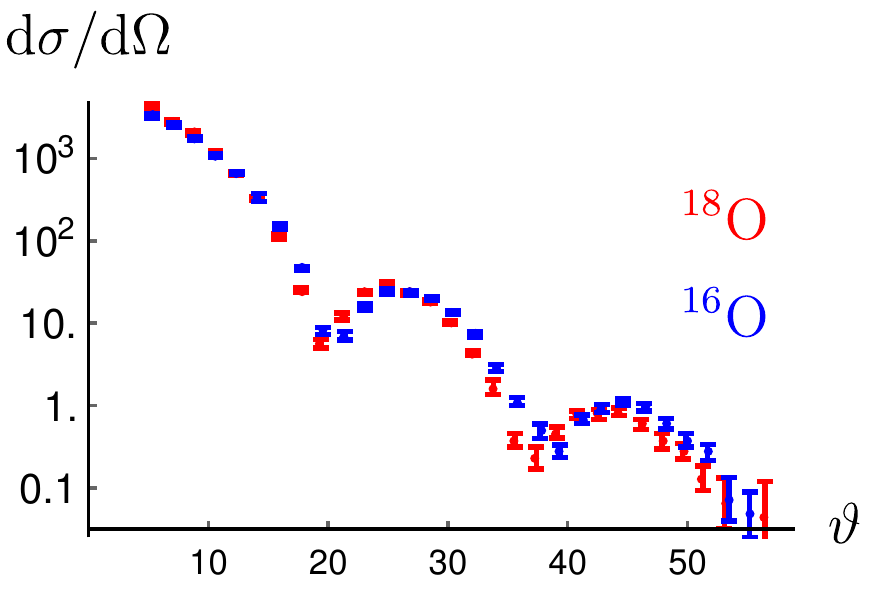}
% antip-16O-178.pdf: 0x0 pixel, 300dpi, 0.00x0.00 cm, bb=
 \caption{Angular distribution for $\bar p$ scattering on \isotope[18]{O} and \isotope[16]{O} at 178.7\,MeV\cite{Bruge:1986fd}.}
 \label{fig:PS184a}
\end{figure}

The results have been analyzed by Lemaire et al.\ in terms of phenomenological optical models \cite{Janouin:1986vh}, which were in turn derived by folding the elementary $\bar N N$ amplitudes with the nuclear density, see, e.g., \cite{Adachi:1987zs,Suzuki:1985xv,Heiselberg:1986vi}.

 In particular, a comparison of \isotope[16]{O} and \isotope[18]{O} isotopes, see Fig.~\ref{fig:PS184a}, reveals that there is very little isospin dependence of the $\bar pN$ interaction, when averaged on spins. 

Other interesting measurements of the antinucleon-nucleus interaction have been carried out and analyzed by the PS179 and OBELIX (PS201) collaborations, with more nuanced conclusions about the isospin dependence of the interaction at very low energy. See, for instance, \cite{Balestra:1988au,Botta:2001fu}.
\subsection{Inelastic scattering}
It has been stressed that the inelastic scattering $\bar p A \to \bar p A^*$, where $A$ is a known excitation of the nucleus $A$, could provide very valuable information on the spin-isospin dependence of the elementary $\bar N N$ amplitude, as the transfer of quantum numbers is identified. One can also envisage the charge-exchange reaction $\bar p A \to \bar n B^{(*)}$. See, for instance, \cite{Dover:1984cq}. 

\begin{figure}[ht!]
\centering
 \includegraphics[width=.4\textwidth]{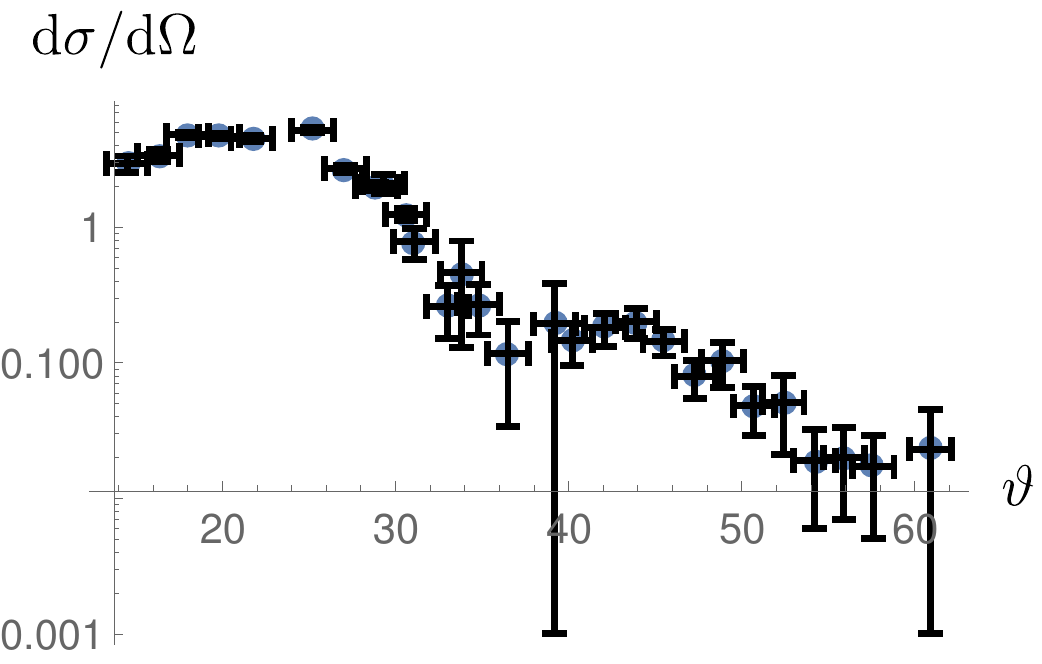}
 % antip-12C3m-179.pdf: 0x0 pixel, 300dpi, 0.00x0.00 cm, bb=
 \caption{Angular distribution of the $\isotope[12]C(\bar p,\bar p)\isotope[12]C^*$ reaction for the $3^-$ excited state at 9.6\,MeV.  The incident $\bar p$ has an energy of 179.7\,MeV}
 \label{fig:12Cexc}
\end{figure}

Some measurements were done by PS184, on $\isotope[12]{C}$ and $\isotope[18]{O}$ \cite{Janouin:1986vh}.  The angular distribution for $\bar p+ \isotope[12]{C}\to \bar p+\isotope[12]{C}^*$ is given in Fig.~\ref{fig:12Cexc} for the case where $\isotope[12]{C}$ is the $3^-$ level at 9.6\,MeV. 
In their analysis, the authors were able to distinguish among models that were equivalent for the $\bar N N$ data, but have some differences in the treatment of the short-range part of the interaction.  This is confirmed by the analysis in \cite{Dover:1984cq,Dover:1984yy}.
Unfortunately, this program of inelastic antiproton-nucleus scattering was not thoroughly carried out. 

\subsection{Antiprotonic atoms}
The physics is nearly the same as for protonium. A low energy antiproton sent towards a target consisting of atoms of nucleus \isotope[A][Z]{X}, is decelerated by the electromagnetic interaction and captured in a high atomic orbit, and cascades down towards lower orbits. During this process, the electrons are expelled.  The difference is that annihilation  occurs before reaching the S  or P levels, actually when the size of the orbit becomes comparable to the size of the nucleus. Again, the Day-Snow-Sucher mechanism can induce some Stark effect.  Thus precocious annihilation can happen, depending on the density of the target.

A review of the experimental data is provided in \cite{Batty:1997zp,Gotta:2004rq}, where a comparison is made with pionic and kaonic atoms. 
The models developed to describe antiproton-nucleus scattering (see Sec.~\ref{subse:pbar-A}) have been applied, and account rather well for the observed shifts. 
As for the purely phenomenological optical potentials $V_\text{opt}$, the most common parametrization is of the form
\begin{equation}\
 2\,\mu\,V_\text{opt}=-4\,\pi\left(1+\frac{\mu}{m}\right)(b_\text{R}+i\,b_\text{I})\varrho(r)~,
\end{equation}
where $\mu$ is the reduced mass of $\bar p$-$A$, $m$ the mass of the nucleon, $\varrho(r)$ the nuclear density and $b_\text{R}+i\,b_\text{i}$ an effective scattering length. 

More refined models,  aiming at describing simultaneously the data on a variety of nuclei, are written as \cite{Batty:1997zp}
% %
% \begin{equation}
%  2\,\mu\,V_\text{opt}=q(r)+ \vec{\nabla}.\alpha(r)\,\vec{\nabla}~,
% \end{equation}
% %
% where $q(r)$ represents a S-wave term and $\vec{\nabla}.\alpha(r)\,\vec{\nabla}$ a P-wave one. They read
% %
\begin{equation}\label{eq:pbar-A-opt}
 2\,\mu\,V_\text{opt}=-4\,\pi\left(1+\frac{\mu}{m}\right)\left(b_0[\varrho_n(r)+\varrho_p(r)]+b_1[\varrho_n(r)-\varrho_p(r)]\right)~,
%  \begin{aligned}
%   q(r)&=-4\,\pi(1+\mu/M)\left(b_0[\varrho_n(r)+\varrho_p(r)]+b_1[\varrho_n(r)-\varrho_p(r)]\right)~,\\
%   \alpha(r)&=4\,\pi(1+\mu/M)^{-1}\,c_0[\varrho_n(r)+\varrho_p(r)]~,
%  \end{aligned}
\end{equation}
where the complex $b_{0,1}$ are the isospin-independent and isospin-dependent effective scattering lengths, respectively.
Further refinements introduce in \eqref{eq:pbar-A-opt} a ``P-wave'' term $\vec{\nabla}.\alpha(r)\,\vec{\nabla}$, or terms proportional to the square of the density. 
% and the $c_0$ the isospin-averaged scattering volume. 
Typical values are \cite{Batty:1997zp}
\begin{equation}
 b_0= (2.7\pm0.3)+ i\,(3.1\pm +0.4)\,\mathrm{fm}~,\qquad
  b_1= (1.2\pm1.6)+ i\,(1.6\pm +1.3)\,\mathrm{fm}~,
\end{equation}
so that there is no firm evidence for a strong isospin dependence. 

It is important to stress that the potential $V_\text{opt}$ is probed mainly at the surface. Its value inside the nucleus hardly matters.  The same property is seen in the low-energy heavy-ion collisions: what is important is the interaction at the point where the two ions come in contact. 
\subsection{Antiproton-nucleus dynamics}
Modeling the antiproton-nucleus interaction has been done with various degrees of sophistication. We have seen in the last section that phenomenological (complex) potentials proportional to the nuclear density account for a wide body of data on antiprotonic atoms. A relativistic mean-field approach was attempted years ago by Bouyssy  and Marcos \cite{Bouyssy:1982cn} and revisited more recently \cite{Gaitanos:2015xpa}. Meanwhile, a Glauber approach has been formulated \cite{Larionov:2016xeb} and applied to the elastic and inelastic scattering of relativistic antiprotons.  

There is a persisting interest in the domain of very low energies and possible bound states. For instance, Friedman et al.~\cite{Friedman:2015qwa} analyzed the subtle interplay between the $\bar NN$ S- and P-waves when constructing the antiproton-nucleus potential.  There has been also speculations about possible $\bar p-A$ states, in line with the studies on the molecular $\bar N N$ baryonium. For a recent update, see, e.g., \cite{HRTANKOVA201845}. 

One could also envisage to use antiprotons to probe the tail of the nuclear density for neutron-rich nuclei with a halo structure. For early refs., see \cite{Bradamante:1990hp}. Recently, the PUMA proposal suggests an investigation by low-energy antiprotons of some unstable isotopes, for which the conventional probes have limitations \cite{Aumann:2691045}. 
\subsection{Neutron-antineutron oscillations}
In some theories of grand unification, the proton decay is suppressed, and one expects neutron-to-antineutron oscillations. An experimental search using free neutrons has been performed at Grenoble \cite{BaldoCeolin:1994jz}, with a limit of about $\tau_{n\bar n} \gtrsim 10^{-8}\,$s for the oscillation period. Any new neutron source motivates new proposals of the same vein, see, e.g.,~\cite{Theroine:2016chp}. 

An alternative is to use the bound neutrons of nuclei. The stability of, say, \isotope[16]{O}, reflects as well the absence of decay of its protons as the lack of $n\to\bar n$ conversion with subsequent annihilation of the antineutron. It has been sometimes argued \cite{Kabir:1983qx} that the phenomenon could be obscured in nuclei by uncontrolled medium corrections.  However, the analysis shows that the neutrons oscillate mainly outside the nucleus, and the subsequent annihilation takes place at the surface, so that, fortunately, the medium corrections are small. 

The peripheral character of the $n\bar n$ oscillations in nuclei explains why  a simple picture  (sometimes called closure approximation) does not work too well, with the neutron and the antineutron in a box feeling an average potential $\langle V_n\rangle$ or $\langle V_{\bar n}\rangle$, resulting in a simple $2\times2$ diagonalization. The true dynamics of $n\bar n$ oscillations relies on the tail of the neutron distribution, where $n$ and $\bar n$ are almost free. 

There are several approaches, see for instance, \cite{Alberico:1998qy}. The simplest is based on the Sternheimer equation, which gives the first order correction to the wave function without summing over unperturbed states.  In a shell model with realistic neutron (reduced) radial  wave functions $u_{n\ell J}(r)$ with shell energy $E_{n\ell J}$, the induced $\bar n$ component is given by
\begin{equation}
 -\frac{w''_{n\ell J}(r)}{\mu}+\frac{\ell(\ell+1)}{\mu\, r^2}+ V_{\bar n}(r)\,w'_{n\ell J}(r) - E_{n\ell J}\,w'_{n\ell J}(r)= \gamma\,u_{n\ell J}(r)~,
\end{equation}
with $\mu$ the reduced mass of the $\bar n$-$(A-1)$ system, $V_{\bar n}$ the complex (optical) $\bar n$-$(A-1)$ potential, and $\gamma=1/\tau_{n\bar n}$ the strength of the transition. Once $w_{n\ell J}$ is calculated, one can estimate the second-order correction to the energy, and in particular the width $\Gamma_{n\ell J}$ of this shell
\begin{equation}\label{eq:deut2}
\Gamma_{n\ell J}=-2\,\int_0^\infty \IM{V_{\bar n}}\,|w_{n\ell J}(r)|^2\,\mathrm{d}r=-2\,\gamma \int_0^\infty u_{n\ell J}(r)\, \IM{w_{n\ell J}(r)}\,\mathrm{d}r~,
\end{equation}
which scales as
\begin{equation}\label{eq:deut3}
 \Gamma_{n\ell J}\propto \gamma^2~.
\end{equation}
An averaging over the shells give a width per neutron $\Gamma$ associated with a lifeftime $T$
\begin{equation}\label{eq:deut4}
T=T_r\,\tau_{n\bar n}^2~,
\end{equation}
where $T_r$ is named either  the ``reduced lifetime'' (in s$^{-1}$) or the ``nuclear suppression factor''. The spatial distribution of the $w_{n\ell J}$ and the integrands in \eqref{eq:deut2}, the relative contribution to $\Gamma$ clearly indicate the peripheral character of the process. See, e.g., \cite{Barrow:2019viz} for an  application to a simulation in the forthcoming DUNE experiment, and refs.\ there to earlier estimates.  Clearly, DUNE will provide the best limit for this phenomenon. 

For the deuteron, an early calculation by Dover et al.~\cite{Dover:1982wv} gave $T_r\simeq 2.5\times 10^{22}\,\mathrm{s}^{-1}$. Oosterhof et al.~\cite{Oosterhof:2019dlo}, in an approach based on effective chiral theory (see Sec.~\ref{se:modern}), found a value significantly smaller, $T_r\simeq 1.1\times 10^{22}\,\mathrm{s}^{-1}$. However, their calculation has been revisited by Haidenbauer and Mei\ss ner \cite{Haidenbauer:2019fyd}, who got almost perfect agreement with Dover et al.  For \isotope[40]{Ar} relevant for the DUNE experiment, the result of  \cite{Barrow:2019viz} is $T_r\simeq 5.6\times 10^{22}\,\mathrm{s}^{-1}$.
\section{Antinucleon annihilation}
\subsection{General considerations}
$N\bar N$ annihilation is a rather fascinating process, in which the baryon and antibaryon structures disappear into mesons. The kinematics is favorable, with an initial center-of-mass energy of  2\,GeV at rest and more in flight, allowing in principle up to more than a dozen of pions. Of course, the low mass of the pion is a special feature of light-quark physics. We notice, however, that the quark model predicts that $(QQQ)+(\bar Q\bar Q\bar Q)>3 (Q\bar Q)$~\cite{Nussinov:1999sx}, so that annihilation at rest remains possible in the limit where all quarks are heavy.  The same quark models suggest that $(\bar Q\bar Q\bar Q)+(qqq)<3\,(\bar Q q)$ if the mass ratio $Q/q$ becomes large, so that, for instance, a triply-charmed antibaryon $(\bar c\bar c\bar c)$ would not annihilate on an ordinary baryon.  

One should acknowledge at the start of this section that there is no theory, nor even any model, that accounts for the many data accumulated on $\bar N N$ annihilation. Actually the literature is scattered across various subtopics, such a the overall strength and range of annihilation, the average multiplicity, the percentage of events with hidden-strangeness, the explanation of specific branching ratios, such as the one for $\bar p p\to \rho\,\pi$, the occurrence of new meson resonances, etc. We shall briefly survey each of these research themes. 
\subsection{Quantum numbers}
%
%A brief reminder is in order. 
An initial $\bar N N$ state with isospin $I$, spin $S$, orbital momentum $L$ and total angular momentum $J$  has parity $P=-(-1)^L$ and $G$-parity $G=(-1)^{I+L+S}$. If the system is neutral, its charge conjugation  is $C=(-1)^{L+S}$.  A summary of the quantum numbers for the $S$ and $P$ states is given in Table~\ref{tab:pw}.
\newcommand{\bs}{\hspace*{-4pt}}
\begin{table}[!h]
\caption{\label{tab:pw}%
Quantum numbers of the S and P partial waves (PW) of the {\NNb} system. The notation is \islj.}
\vskip 4pt
\renewcommand{\arraystretch}{1.2}
\begin{center}
\begin{tabular}{ccccccccccccc}
\hline\hline
\bs PW \bs&\bs\ISLJ11S0\bs&\bs\ISLJ31S0\bs&\bs\ISLJ13S1\bs&\bs\ISLJ33S1\bs&%
\bs\ISLJ11P1\bs&\bs\ISLJ31P1\bs&\bs\ISLJ13P0\bs&\bs\ISLJ33P0\bs&%
\bs\ISLJ13P1\bs&\bs\ISLJ33P1\bs&\bs\ISLJ13P2\bs&\bs\ISLJ33P2\bs\\
\hline
$J^{PC}$  &$ 0^{-+}   $&$ 0^{-+}   $&$ 1^{--}  $&$ 1^{--}    $&$% singlet and triplet S
 1^{+-}   $&$ 1^{+-}   $&$ 0^{++}   $&$  0^{++}  $&$% 1P1 and 3P0
 1^{++}   $&$  1^{++}  $&$ 2^{++}   $&$  2^{++}  $\\% 3P1 and 3P2 
$I^G$ &$ 0^+   $&$ 1^-  $&$  0^- $&$  1^+   $&$% singlet and triplet S
 0^-   $&$ 1^+   $&$ 0^+  $&$  1^-  $&$% 1P1 and 3P0
 0^+   $&$ 1^-   $&$ 0^+  $&$   1^- $\\% 3P1 and 3P2 
\hline\hline\end{tabular} \end{center}    
\end{table}

So, for a given initial state, some transitions are forbidden or allowed. The result for some important channels is shown in Table~\ref{tab:ann:allowed}. In particular, producing two identical scalars or pseudoscalars requires an initial P-state. 
\renewcommand{\bs}{\hspace*{-3pt}}
\renewcommand{\baselinestretch}{1.3}
\newcommand{\y}{$\surd$}
\begin{table}[!h]
\caption{\label{tab:ann:allowed}%
Allowed decays from S and P-states into some two-meson final states.}
\begin{center}\begin{tabular}{ccccccccccccc}
\hline\hline
\bs FS\bs&\bs\ISLJ11S0\bs&\bs\ISLJ31S0\bs&\bs\ISLJ13S1\bs&\bs\ISLJ33S1\bs&%
\bs\ISLJ11P1\bs&\bs\ISLJ31P1\bs&\bs\ISLJ13P0\bs&\bs\ISLJ33P0\bs&%
\bs\ISLJ13P1\bs&\bs\ISLJ33P1\bs&\bs\ISLJ13P2\bs&\bs\ISLJ33P2\bs\\
\hline
$\piz\piz$&     &       &     &     &%S-states
    &     & \y    &      &% 1P1 and 3P0
    &    &   \y &       \\%  3P1 and 3P2
 $\pim\pip$&     &       &     &   \y  &%S-states
    &     & \y    &      &% 1P1 and 3P0
    &    &   \y &       \\%  3P1 and 3P2 
 $\piz\eta^{(\prime)}$&     &       &     &     &%S-states
    &     &    &  \y     &% 1P1 and 3P0
    &    &   &   \y     \\%  3P1 and 3P2    
 $\eta^{(\prime)}\eta^{(\prime)} $&     &       &     &     &%S-states
    &     &  \y  &       &% 1P1 and 3P0
    &    &   \y &        \\%  3P1 and 3P2   
 $\Km\Kp$&     &       & \y    &   \y  &%S-states
    &     & \y   &  \y     &% 1P1 and 3P0
    &    &\y   &   \y     \\%  3P1 and 3P2   
 $\Ks\Kl$&     &       & \y    &   \y  &%S-states
    &     &   &       &% 1P1 and 3P0
    &    &   &      \\%  3P1 and 3P2   
 $\Ks\Ks$&     &       &     &   &%S-states
    &     & \y  &  \y     &% 1P1 and 3P0
    &    &  \y & \y     \\%  3P1 and 3P2    
$\piz\omega(\phi)$&     &       &     &\y   &%S-states
    &  \y   &   &       &% 1P1 and 3P0
    &    &   &       \\%  3P1 and 3P2   
\bs $\eta^{(\prime)}\omega(\phi)$\bs&     &       & \y    &   &%S-states
 \y   &     &    &       &% 1P1 and 3P0
    &    &   &     \\%  3P1 and 3P2 
$\piz\rho^0$&     &       & \y    &   &%S-states
 \y   &     &    &       &% 1P1 and 3P0
    &    &   &     \\%  3P1 and 3P2    
$\eta^{(\prime)}\rho^0$&     &       &    & \y   &%S-states
    &   \y   &    &       &% 1P1 and 3P0
    &    &   &     \\%  3P1 and 3P2    
 $\pi^\pm\rho^\mp$&     &    \y   & \y   &   &%S-states
 \y   &      &    &       &% 1P1 and 3P0
    &  \y  &   &   \y  \\%  3P1 and 3P2              
\hline\hline
\end{tabular}  \end{center}
\renewcommand{\baselinestretch}{1.0}
\end{table}

The algebra of quantum numbers is not always trivial, especially if identical mesons are produced. For instance, the question was raised whether or not the $\SLJ1S0$ state of protonium with $J^{PC}=0^{-+}$ and $I^G=0$ can lead to a final state made of four $\pi^0$.  An  poll among colleagues gave an overwhelming majority of negative answers. But a transition such as $\SLJ1S0\to 4\,\pi^0$ is actually possible at the expense of several internal orbital excitations among the pions. For an elementary proof,  see \cite{Klempt:2005pp}, for a more mathematical analysis~\cite{Zupancic:1994bf}.

The best known case, already mentioned in Sec.~\ref{se:disco}, deals with $\pi\pi$. An $S$-wave $\pi^+\pi^-$ with a flat distribution, or a $\pi^0\pi^0$ system (necessarily with $I=0$ and $J$ even) requires an initial state $\ISLJ13P0$. It has been observed to occur even in annihilation at rest on a dilute hydrogen target~\cite{Devons:1971rn}. This is confirmed by a study of the $J=0$ vs.\ $J=1$ content of the $\pi\pi$ final state as a function of the density of the target, as already mentioned in Sec.~\ref{subse:DSN}.

% A word here about the Day-Snow-Sucher effect (DSS) that plays a role in understanding why annihilation at rest depends on the density. (\emph{Perhaps to be moved in the section about protonium}). A low-energy  antiproton sent on an hydrogen target is further slowed down by the it electromagnetic interaction with the medium and then captured by an atom in a rather high orbit, which is such that the radius of the antiproton orbit is of the order of the size of the electronic orbits.  The electron is expelled and the antiproton cascades down using preferentially circular orbits with principal number $n=L$. If the protonium is isolated its annihilation probability is negligible for $L\ge 2$, or the order of $1\,\%$ for $L=1$, so that annihilation occurs mainly for $L=0$, tough $L=1$ annihilation can be detected. In a target, the compact protonium often travels inside the orbits of an ordinary atom, where it experiences a strong electric field. The Stark effect populates $nS$ which annihilate. This shortens the lifetime of the protonium, as the electromagnetic cascade is interrupted and modify the ratio of $S$-to-$P$ annihilations. 
%
\subsection{Global picture of annihilation}
\begin{figure}[ht!]
 \centerline{
 \includegraphics[width=.20\textwidth]{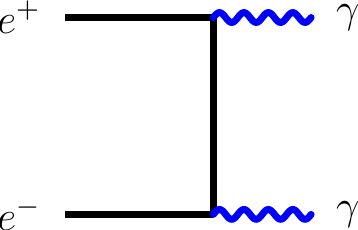}
 \qquad
  \includegraphics[width=.20\textwidth]{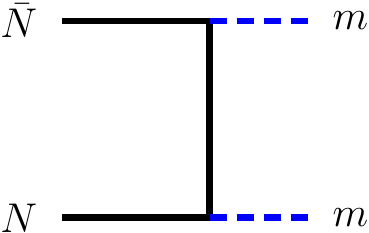} }
 % figsreview-fig22.pdf: 0x0 pixel, 300dpi, 0.00x0.00 cm, bb=
 \caption{Left: $e^+e^-$ annihilation into two photons. Right: $\bar N N$ annihilation into two mesons}
 \label{fig:ann-epem}
\end{figure}

\begin{figure}[ht!]
 \centerline{
 \includegraphics[width=.30\textwidth]{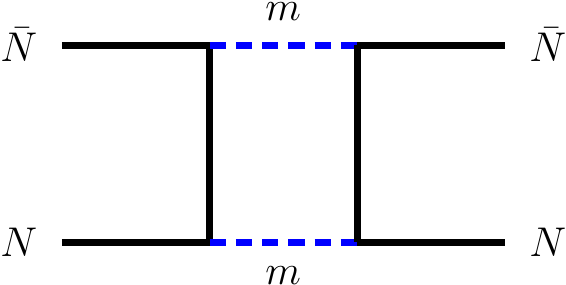}}
 \caption{Typical contribution to the absorptive part of the $\bar N N$ amplitude }
 \label{fig:ann-ite}
\end{figure}

As already stressed, the main feature of annihilation  is its large cross-section, which comes together  with a suppression of the charge-exchange process.
This is reinforced by the observation that even at rest,  annihilation is not reduced to an S-wave phenomenon. 
% It was, indeed,  rather a surprise to detect final states that require an initial P-wave, such as  $\bar p p\to \pi^0 \pi^0$ \cite{Devons:1971rn}.
% %\cite{Pinski:1960xga,Erwin:1961ny,Chadwick:1962sja}
%
This is hardly accommodated with a zero-range mechanism such as baryon exchange.  The baryon exchange, for say, annihilation into two mesons  is directly inspired  by electron exchange in $e^+\,e^-\to \gamma\,\gamma$. See Fig.~\ref{fig:ann-epem}.
After iteration, the absorptive part of the $\bar N N$ interaction, in this old-fashioned picture, would be driven by diagrams such as the one in Fig.~\ref{fig:ann-ite}.
Other contributions involve more than two mesons and crossed diagrams. As analyzed, e.g., in \cite{PhysRevLett.5.380,PhysRev.124.614}, this corresponds to  a very small range, practically a contact interaction.  Not surprisingly, it was impossible to follow this prescription when building optical models to fit the observed cross-sections. Among the contributions, one may cite \cite{1958NCim....8..485G,Bryan:1968ns,PhysRevC.21.1466,Kohno:1986fk}. Claims such as \cite{PhysRevLett.48.1319}, that it is possible to  fit the cross sections  with a short-range annihilation operator,  are somewhat flawed by the use of  very large strengths, wide form factors, and a momentum dependence of the optical potential that reinforce annihilation in $L>0$ partial waves.

In the 80s, another point of view started to prevail:  annihilation should be understood at the quark level.%
\footnote{Of course , the rearrangement was introduced much earlier, in particular by Stern, Rubinstein, Caroll,  \dots \cite{Rubinstein:1966zza,Carroll:1970ww}, but simply to calculate ratios of branching ratios, without any attempt to estimate the cross sections.}\@
This picture was hardly accepted by a fraction of the community. 
An anecdote illustrates how hot was the debate. After a talk at the 1988 Mainz conference on antiproton physics, where I presented the quark rearrangement, Shapiro strongly objected. At this time, the questions and answers were recorded and printed in the proceedings. Here is the verbatim \cite{Kleinknecht:1989ep}:  {\sl \textit{I.S. Shapiro} (Lebedev) : The value of the annihilation range \dots is not a question for discussion. It is a general
statement following from the analytical properties of the amplitudes in quantum field theory \dots. It does not matter how the annihilating objects are constructed from
their constituents. It is only important that, in the scattering induced by annihilation, an energy of at
least two baryons masses is transferred. 
\textit{J.M. Richard}: First of all, for me, this is an important "question for discussion". In fact, we agree
completely in the case of "total annihilation", for instance $\bar N N\to \phi\phi$. The important point is that
[baryons and] mesons are composite, so, what we call "annihilation" is, in most cases, nothing but a soft
rearrangement of the constituents, which does not have to be short range.}

\begin{figure}[ht!]
 \centering
 \includegraphics[width=.35\textwidth]{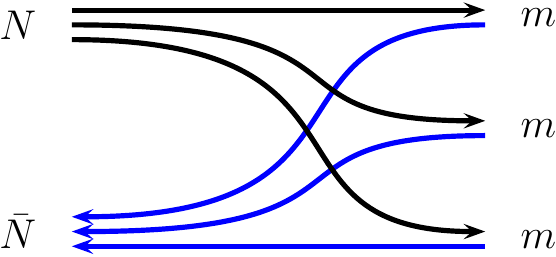}
 \caption{Rearrangement of the quarks and antiquarks from a baryon and an antibaryon  to a set of three mesons}
 \label{fig:rearr}
\end{figure}

In the simplest quark scenario, the spatial dependence of ``annihilation'' comes from that this is not an actual annihilation similar to $e^+e^- \to \text{photons}$, in which the initial constituents disappear, but a mere \emph{rearrangement} of the quarks, similar  to the rearrangement of the atoms  in some molecular collisions.  This corresponds to the diagram of Fig.~\ref{fig:rearr}.
The amplitude for this process is $\langle\Psi_f|H|\Psi_i\rangle$, where $H$ is the 6-quark Hamiltonian, $\Psi_i$ the nucleon-antinucleon initial state, and $\Psi_f$ the final state made of three mesons. See, for instance, \cite{Green:1985yy,Green:1985vu,Ihle:1988mp}, for the details about the formalism. One gets already a good insight on the spatial distribution of annihilation within the quark-rearrangement model by considering the mere overlap $\langle\Psi_f|\Psi_i\rangle$ using simple oscillator wave functions.  For the initial state, a set of Jacobi coordinates (here normalized to correspond to an unitary transformation)
\begin{equation}
 \label{eq:Jacobi-B}
 \begin{aligned}
 \vec x&=\frac{\vec r_2-\vec r_1}{\sqrt2}~,\quad &\vec y&=\frac{2\,\vec r_3-\vec r_1-\vec r_2}{\sqrt6}~,\quad & 
 \vec z&=\frac{\vec r_1+\vec r_2+\vec r_3}{\sqrt3}~,
 \\
\vec x'&=\frac{\vec r_5-\vec r_4}{\sqrt2}~, \quad&\vec y'&=\frac{2\,\vec r_6-\vec r_4-\vec r_5}{\sqrt6}~,\quad & 
\vec z'&=\frac{\vec r_4+\vec r_5+\vec r_6}{\sqrt3}~,
\end{aligned}
\end{equation}
with the further change
\begin{equation}
 \label{eq:Jacobi-B1}
 \vec r=\frac{\vec z'-\vec z}{\sqrt2}~,\qquad \vec R=\frac{\vec{z}+\vec z'}{\sqrt2}~,
\end{equation}
which, to a factor, are the  $\bar N N$ separation and the overall center of mass.
The initial-state wave function is thus of the form
\begin{equation}
 \label{eq:psii} \Psi_i=\genfrac{(}{)}{}{}{\alpha}{\pi}^{\!\!3}\,\exp\left[-(\alpha/2)(\vec x^2+\vec y^2+\vec x'^2+\vec y'^2)\right]\,\varphi(\vec r)~,
\end{equation}
where $\varphi$ denotes the $\bar N N$ wave function. Similarly for the final state, one can introduce the normalized Jacobi coordinates
\begin{equation}
 \label{eq:Jacobi-m}
 \begin{aligned}
 \vec u_1&=\frac{\vec r_4-\vec r_1}{\sqrt2}~,\quad &\vec u_2&=\frac{\vec r_5-\vec r_2}{\sqrt2}~,\quad & 
 \vec u_3&=\frac{\vec r_6-\vec r_3}{\sqrt2}~,
 \\
\vec v_1&=\frac{\vec r_1+\vec r_4}{\sqrt2}~,\quad &\vec v_2&=\frac{\vec r_2+\vec r_5}{\sqrt2}~,\quad & 
 \vec v_3&=\frac{\vec r_3+\vec r_6}{\sqrt2}~,\\
 \vec X&=\frac{\vec v_2-\vec v_1}{\sqrt2}~,\quad &  \vec Y&=\frac{2\,\vec v_3-\vec v_1-\vec v_2}{\sqrt6}~,\quad
& \vec R&=\frac{\vec v_1+\vec v_2+\vec v_3}{\sqrt3}~,
\end{aligned}
\end{equation}
and the wave function
\begin{equation}
\Psi_f=\genfrac{(}{)}{}{}{\beta}{\pi}^{\!\!9/4}\,\exp\left[-(\beta/2)(\vec u_1^2+\vec u_2^2+\vec u_3^2)\right]\,\Phi(\vec X,\vec Y)~.
\end{equation}
Integrating for instance over $\vec x'-\vec x$ and $\vec y'-\vec y$, one ends with 
\begin{equation}
 \Phi_f^*(\vec X,\vec Y)\, \exp(-\beta \vec r^2/2)\,\exp(-\alpha(\vec X^2+\vec Y^2)/2)\, \varphi(\vec r)~,
\end{equation}
and after iteration, one gets a separable operator $v(r)\,v(r')$, where  $v(r)$ is proportional to $\exp(-\beta \vec r^2/2)$ and contains some energy-dependent factors \cite{Green:1985yy,Ihle:1988mp}.  As expected, the operator is not local. There is an amazing exchange of roles: the size of the baryon, through the parameter $\alpha$, governs the spatial spread of the three mesons, while the size the mesons becomes the range of the separable potential. Schematically speaking, the range of ``annihilation'' comes from the ability of the mesons to make a bridge, to pick up a quark in the baryon and an antiquark in the antibaryon. 

Explicit calculations show that the rearrangement potential has about the required strength to account for the observed annihilation cross-sections. Of course, the model should be improved to include the unavoidable distortion of the initial- and final-state hadrons. Also one needs a certain amount of intrinsic quark-antiquark annihilation and creation to explain the production of strange mesons.  This leads us to the discussion about the branching ratios.
\subsection{Branching ratios: experimental results}
Dozens of final states are available for $\bar N N$ annihilation, even at rest. When the energy increases, some new channels become open. For instance, the $\phi\phi$ channel was used to search for glueballs in the PS202 experiment \cite{Eisenstein:1997cn}. However, most measurements have been performed at rest with essentially two complementary motivations. The first one was to detect  new multi-pion resonances, and, indeed, several mesons have been either discovered or confirmed thanks to the antiproton-induced reactions. The second motivation was to identify some leading mechanisms for annihilation, and one should confess that the state of the art is not yet very convincing. 

Several reviews contain a summary of the available branching ratios and a  discussion on their interpretation. See, e.g., \cite{Amsler:1991cf,Klempt:2005pp}. We shall not list all available results, but, instead, restrict ourselves to the main features or focus on some intriguing details. For instance:
\begin{itemize}
 \item 
 The average multiplicity is about 4 or 5. But in many cases, there is a formation of meson resonances, with their subsequent decay.  In a rough survey, one can estimate  that a very large fraction of the annihilation channels are compatible with the primary formation of two mesons which subsequently decay. 
 \item
 In the case of a narrow resonance, one can distinguish the formation of a resonance from a background made of uncorrelated pions, e.g., $\omega\pi$ from $\pi\pi\pi\pi$. In the case of broad resonances, e.g., $\rho\pi$ vs.\ $\pi\pi\pi$, this is much more ambiguous. 
 \item The amount of strangeness, in channels such as $\bar K+K$, $\bar K+ K^*$, $\bar K +K+\text{pions}$, is about 5\%.
 \item Charged states such as $\bar p n$ or $\bar n p$ are pure isospin $I=1$ initial state. In the case of $\bar p p$ annihilation at rest, the isospin is not known, except if deduced from the final state, like in the case of $\pi\eta$. Indeed, $\bar p p$ is the combination $(|I=0\rangle+|I=1\rangle)/\sqrt2$. But, at short distances, one of the components often prevails, at least in model calculations. In the particle basis, there is an admixture of $\bar n n$ component, which, depending on its relative sign, tends to make either a dominant $I=0$, or $I=1$.  For instance,  Kudryavtsev \cite{Kudryavtsev:1999hj}  analyzed the channels involving two pseudoscalars, and concluded that if protonium annihilation is assumed to originate from an equal mixture of $I=0$ and $I=1$, then annihilation is suppressed in one of the isospin channels, while a better understanding is achieved, if $\bar p p-\bar n n$ is accounted for. 
\end{itemize}
\subsection{Branching ratios: phenomenology}
The simplest model, and most admired, is due to Vandermeulen \cite{Vandermeulen:1988hh}. It assumes a dominance of 2-body modes, say $\bar N N\to a+b$, where $a$ and $b$ are mesons or meson resonances, produced preferentially when the energy is slightly above the threshold $s_{ab}^{1/2}=m_a+m_b$. More precisely, the branching ratios are parametrized as
\begin{equation}
 f=C_{ab}\,p(\sqrt{s}, m_a, m_b)\,\exp[-A(s-s_{ab})]~,
\end{equation}
where $A$ is an universal parameter, $p$ the center-of-mass momentum and the constant $C_{ab}$ assume only two values: $C_0$ for non-strange and $C_1$ for strange. 

In the late 80s, following the work by Green and Niskanen \cite{Green:1985yy,Green:1985vu}, and others, there were attempts to provide a detailed picture of the branching ratios, using quark-model wave functions supplemented by operators to create or annihilate quark-antiquark pairs. A precursor was the so-called $\SLJ3P0$ model \cite{LeYaouanc:111431} introduced to describe decays such as $\Delta\to N+\pi$.

\begin{figure}[ht!]
 \centering
 \includegraphics[width=.4\textwidth]{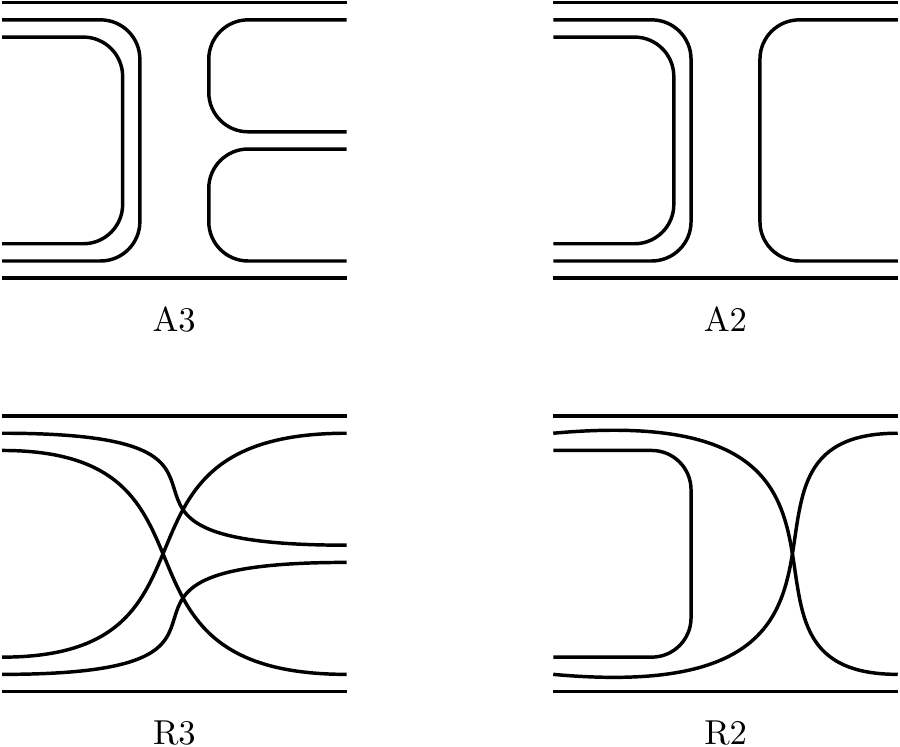}
 % Diags-fig1.pdf: 0x0 pixel, 300dpi, 0.00x0.00 cm, bb=
 \caption{Some quark diagrams describing annihilation}
 \label{fig:quark-diag}
\end{figure}

There has been attempts to understand the systematics of branching ratios at the quark level. We already mentioned some  early papers \cite{Rubinstein:1966zza,Carroll:1970ww}. In the late 80s and in the 90s, several papers were published, based on a zoo of quark diagrams. Some of them are reproduced in Fig.~\ref{fig:quark-diag}.
The terminology adopted A$n$ or R$n$ for annihilation or rearrangement into $n$ mesons. Of course, they are not Feynman diagrams, but just a guidance for a quark model calculation with several assumptions to be specified.  On the one had, the R3 diagram comes as the most ``natural'', as it does not involve any change of the constituents. On the other hand, it was often advocated that planar diagrams should be dominant, see, e.g., \cite{Genz:1988fz}. This opinion was, however, challenged by Pirner in his re-analysis the the $1/N_c$ expansion, where $N_c$ is the number of colors in QCD~ \cite{Pirner:1988mn}.

A key point is of course strangeness. The R3 diagram hardly produces kaons, except if extended as to include the sea quarks and antiquarks. On the other hand, the A$n$ diagrams tend to produce too often kaons, unless a controversial \emph{strangeness suppression factor} is applied: at the vertex where a quark-antiquark pair  is created, a factor $f=1$ is applied for $q=u,\,d$ and $f\ll 1$ for $q=s$.  This is an offending violation of the flavor SU(3)$_F$ symmetry.  For instance the decays $J/\psi\to p\bar p$ and $J/\psi\to\Lambda\bar\Lambda$ are nearly identical, especially once phase-space corrections are applied.  The truth  is that at low-energy, strangeness is dynamically suppressed by phase-space and a kind of tunneling effect~\cite{Dosch:1986sf}.  This could have been implemented more properly in the analyses of the branching ratios. An energy-independent strangeness suppression factor is probably too crude. 

Note that a simple phenomenology of quark diagrams is probably elusive. A diagram involving two primary mesons can lead to 4 or 5 pions after rescattering or the decay of a resonance.  Also the A$n$ diagrams require a better overlap of the initial baryon and antibaryon, and thus are of shorter range than the R$n$ diagrams. So the relative importance can vary with the impact parameter and the incident energy. 
\subsection{Annihilation on nuclei}
There has been several studies of $\bar N$-$A$ annihilation. In a typical scenario, a primary annihilation produces mesons, and some of them penetrate the nucleus, giving rise to a variety of phenomenons: pion production, nucleon emission, internal excitation, etc. See, e.g., \cite{Cugnon:1996xt}. Some detailed properties have been studied, for instance whether annihilation on nuclei produces less or more strange particles than annihilation on nucleons~\cite{Cugnon:1991nd}.

At very low energy, due to the large $\bar NN$ cross section, the primary annihilation takes place near the surface. It has been speculated that with medium-energy antiprotons, thanks to the larger momentum and the smaller cross section, the annihilation could sometimes take place  near the center of the nucleus. Such rare annihilations with a high energy release (at least 2\,GeV) and little pressure, would explore a sector of the properties of the nuclear medium somewhat complementary to the heavy-ion collisions.  See, e.g., \cite{Dalpiaz:1987qk,Rafelski:1996ik,Koch:2004db,Wiedner:2009uwz}.

Note the study of Pontecorvo reactions. In $\bar N N$ annihilation, at least two mesons have to be produced, to conserve energy and momentum. On a nucleus, there is the possibility of primary annihilation into $n$ mesons, with $n-1$ of them being absorbed by the remaining nucleons. An example is $\bar p NN\to \pi N$ or $\phi n$  \cite{Chiba:1996tf,Gorchakov:2002dm}.  This is somewhat  related to the pionless decay of $\Lambda$ in hypernuclei \cite{Alberico:2001jb}.
\subsection{Remarkable channels}
Some annihilation channels have retained the attention:
\begin{itemize}
 \item $p\bar p\to e^+e^-$ led to a measurement of the proton form factor in the time-like region. The reversed reaction $e^+e^-\to p\bar p$ was studied elsewhere, in particular at Frascati. For a general overview, see \cite{Denig:2012by,Pacetti:2015iqa}, and for the results of the PS170 collaboration at CERN, \cite{Bardin:1991rz}.
 \item We already mentioned the $\bar p p\to \text{charmonium} \to \text{hadrons}$, leading to a better measurement of the width of some charmonium states, and the first indication for the $h_c$, the $\SLJ1P1$ level of $c\bar c$ \cite{Baglin:1986yd,Baglin:1985xa}. In principle, while $e^+e^-\to\text{charmonium}$ is restricted to the $J^{PC}=1^{--}$ states, $\bar p p$ can match any partial wave. However, perturbative QCD suggests that the production is suppressed for some quantum numbers. It was thus a good surprise that $\eta_c(1S)$ was seen in $\bar p p$, but the coupling turns out less favorable for $\eta_c(2S)$ \cite{Ambrogiani:2001wg,Patrignani:2004nf}.
 \item The overall amount of hidden-strangeness is about $5\,\%$ \cite{Klempt:2005pp}. This is remarkably small and is hardly accommodated in models where several incoming $q\bar q$ pairs are annihilated and several quark-antiquark pairs created.  Note that the branching ratio for $K^+K^-$ is significantly larger for an initial S-wave than for P-wave \cite{Amsler:2019ytk}. This confirms the idea that annihilation diagrams are of shorter range than than the rearrangement ones. 
  \item $\bar p p\to K^0 \bar K{}^0$ in the so-called CPLEAR experiment (PS195) \cite{Gabathuler:2004qm} gave the opportunity to measure new parameters of the $CP$ violation in the neutral $K$ systems, a phenomenon first discovered at BNL in 1964 by Christenson, Cronin, Fitch  and Turlay.\footnote{Happy BNL director, as his laboratory also hosted the experiment in which the $\Omega^-$ was discovered, the same year 1964.} The CPLEAR experiment found evidence for a direct $T$-violation (time reversal). 
 \item Precision measurements of the $\bar p p\to \gamma+X$ and $\bar p p\to \pi+X$ in search of bound baryonium, of which some indications were found before LEAR. The results of more intensive searches at LEAR were unfortunately negative. See, e.g., \cite{Hatzifotiadou:1986ax}.  When combined to the negative results of the scattering experiments, this was seen as the death sentence of baryonium.  But, as mentioned in Sec.~\ref{se:baryonium}, this opinion is now more mitigated, because of the $p\bar p$ enhancements observed in the decay of heavy particles. 
 \item $\bar p p\to \rho\,\pi$ has intriguing properties. Amazingly, the same decay channel is also puzzling in charmonium decay, as the ratio of $\psi(2S)\to  \rho\,\pi$ to $J/\psi\to \rho\,\pi$ differs significantly from its value for the other channels. See, e.g., \cite{Wang:2012mf} and refs.\ there. In the case of $\bar p p$ annihilation, the problem, see, e.g., \cite{Amsler:2019ytk}, is that the the production from $\ISLJ13S1$ is much larger than from $\ISLJ13S0$.  Dover et al., for instance, concluded to the dominance of the A2 type of diagram \cite{Dover:1986jh}, once the quark-antiquark creation operator is assumed to be given by  the $\SLJ3P0$ model \cite{LeYaouanc:111431}.  But the A2 diagram tends to produce too often kaons!
 \item $\bar p N\to \bar K+X$, if occurring in a nucleus, monitors the production of heavy hypernuclei. It was a remarkable achievement of the LEAR experiment PS177 by Polikanov et al.  to measure the lifetime of heavy hypernuclei. See, e.g., \cite{Nifenecker:1992uz}.
 \item $p\bar p\to \pi^+\pi^-$ and $p \bar p\to K^+K^-$ by PS172 revealed striking spin effects (see Sec.~\ref{subse:ann:2ps})
 \item $p\bar p\to \phi\phi$ was used to  search for glueballs (Experiment PS 202 ``JETSET'') \cite{Eisenstein:1997cn}, with innovative detection techniques. 
\end{itemize}
\section{Modern perspectives}\label{se:modern}
So far in this review, the phenomenological interpretation was based either on the conventional meson-exchange picture or on the quark model for annihilation.  The former was initiated in the 50s, and the latter in the 80s. Of course, it is not fully satisfactory to combine two different pictures, one for the short-range part, and another for the long-range, as the results are very sensitive to the assumptions for the matching of the two schemes.  This is one of the many reasons why the quark-model description of the short-range nucleon-nucleon interaction has been abandoned, though it provided an interesting key for a simultaneous calculation of all baryon-baryon potentials. 
One way out that was explored consists of exchanging the mesons between quarks. Then the quark wave function generates a form factor.  For $N\bar N$, a attempt was made by Entem and Fern\'andez~\cite{Entem:2006dt}, 
with some phenomenological applications. In this paper, the annihilation potential is   
due to transition $q\bar q\to\hbox{meson}\to q\bar q$ or $q\bar q\to \hbox{gluon}\to q\bar q$. 
But this remains a rather hybrid picture and  it was not further developed. 

Somewhat earlier, in the 80s, interesting developments of the bag model have been proposed, where the nucleon is given a pion cloud that restores its interaction with other nucleons. This led to a solitonic picture, e.g.,  Skyrme-type of models  for low-energy hadron physics \cite{Zahed:1986qz}.  A first application to $\bar N N$ was proposed by Zahed and Brown \cite{Zahed:1987us}.

As seen in other chapters of this book, a real breakthrough was provided by the advent of effective chiral theories, with many successes, for instance in the description of the  $\pi\pi$ interaction. For a general introduction, see, e.g., the textbook by Donnelly et al.~\cite{Donnelly:2017aaa}.  This approach was adopted by a large fraction of the nuclear-physics community, and, in particular, it was applied to the study of nuclear forces and nuclear structures. 
Chiral effective field theory led to very realistic potentials for the $NN$ interaction, including the three-body forces and higher corrections in a consistent manner  \cite{Machleidt:2011zz,Epelbaum:2014efa}. Thus the meson-exchange have been gradually forsaken. 

\begin{figure}[ht!]
\centering
\includegraphics[width=.5\textwidth]{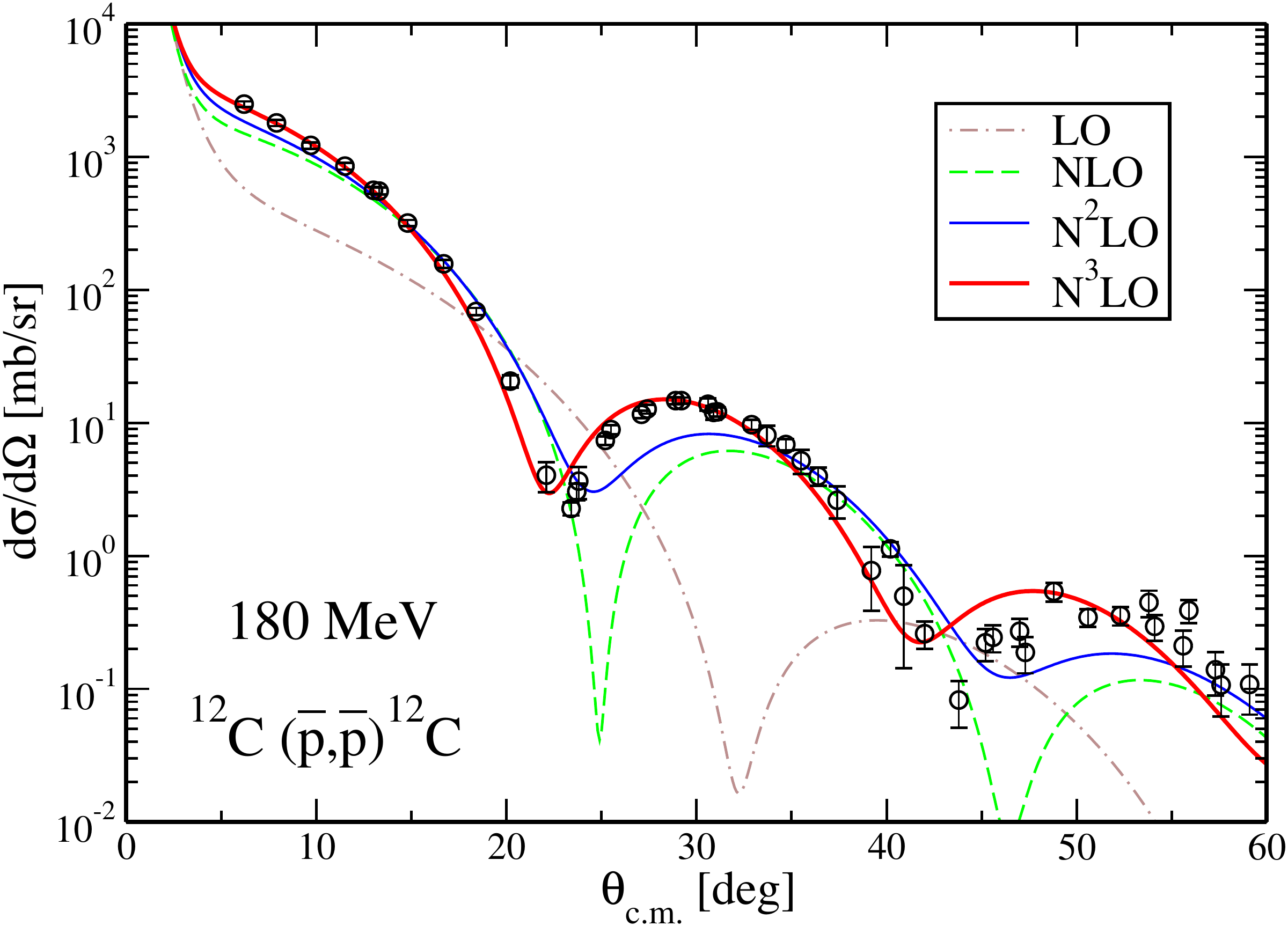}
 % pbar_12C_180MeV_orders.pdf: 0x0 pixel, 300dpi, 0.00x0.00 cm, bb=
 \caption{Differential cross-section for elastic  $\bar p$ scattering on \isotope[12]{C} at 180\,MeV. The optical potential is computed from successive refinements in the effective theory  \cite{Vorabbi:2019ciy}}
 \label{fig:12C-chiral}
\end{figure}

In such modern $NN$ potentials, one can identify the long-range part due to one-, two- or three-pion exchange, and apply the $G$-parity rule, to derive the corresponding long-range part of the  $\bar N N$ potential. The short-range part of the $NN$ interaction is determined empirically,  by fixing the strength of a some constant terms which enter the interaction in this approach. This part cannot  be translated as such to the $\bar N N$ sector.  
There exists for sure, analogous constant terms that describe the real part of the interaction.   As for the annihilation part, there are two options. The first one consists of making the contact terms complex.  This is the choice made by Chen et al;~\cite{Chen:2011yu}. Another option that keeps unitarity more explicit  is to introduce a few effective meson channels $X_i$ and iterate, i.e., $\bar N N \to X \to \bar N N$, with the propagator of the mesonic channel $X_i$ properly inserted \cite{Dai:2017ont}. Then some empirical constant terms enter now the transition potential $V(\bar N N\to X_i)$. A fit of the available data determines in principle the constants of the model \cite{Zhou:2012ui}.  The question remains whether the fit of the constant terms is unique, given the sparsity of spin observables.  For a recent review on chiral effective theories applied to antiproton physics, see \cite{Haidenbauer:2018wso,Haidenbauer:2019ect}. 
The phenomenology will certainly extent beyond scattering data. 
One can already notice that  the amplitude of \cite{Dai:2017ont}, when properly folded with the nuclear density, provides with an optical potential that accounts fairly well for the scattering data, as seen in Fig.~\ref{fig:12C-chiral} borrowed from \cite{Vorabbi:2019ciy}.
\section{Outlook}
The physics of low-energy antiprotons covers a variety of topics: fundamental symmetries, atomic physics, inter-hadronic  forces, annihilation mechanisms, nuclear physics, etc. 

New experiments are welcome or even needed to refine our understanding of this physics. For instance,  a better measurement of the shift and width of the antiprotonic  lines, and some more experiments on  the scattering of antineutrons off nucleons or nuclei. We also insisted on the need for more measurements on $\bar p p$ scattering with a longitudinally or transversally polarized target.

Selected annihilation measurements could also be useful, from zero energy to well above the charm threshold, and again, the interest is twofold: access to new sectors of hadron spectroscopy, and test the mechanisms of annihilation.  For this latter purpose,  a through comparison of $\bar N N$- and $\bar Y N$-induced channels would be most useful, where $Y$ denotes a hyperon. 

The hottest sectors remain these linked to astrophysics: how antiprotons and light antinuclei are produced in high-energy cosmic ray? Is there a possibility in the early Universe of separating matter from antimatter before complete annihilation? Studying these questions require beforehand a good understanding of the antinucleon-nucleon and antinucleon-nucleus interaction. 
\subsection*{Acknowledgments} I would like to thank many colleagues for several stimulating and entertaining discussions along the years about the physics of antiprotons, and in particular, my collaborators.  I would like also to thanks the editors of Frontiers in Physics for initiating this review, and their referees for constructive comments. Special acknowledgments  are due to M.~Asghar for remarks and suggestions. Toto

%
% \bibliographystyle{unsrt}
% %\bibliography{utphys}
% \bibliography{newbib}
% \end{document}
%

\end{document}